\documentclass[iop]{emulateapj}
\usepackage{amsmath} 
\usepackage[utf8x]{inputenc}
\usepackage{natbib}
\usepackage{amssymb}
\usepackage{xfrac}

\usepackage{hyperref}
\usepackage{color}
\definecolor{MidnightBlue}	{RGB}{ 25,  25, 112}
\definecolor{Navy}		{RGB}{  0,   0, 128}
\hypersetup{colorlinks=true, linkcolor=Navy, citecolor=Navy, urlcolor=Navy}

\def \yr        {{\rm\,yr}}

\def \kev       {{\rm\ keV}}

\def \logTd6 {\hbox{log$( T/6 \kev)$} }

\def \Mag	{\,\mathrm{mag}}		

\def \Ry	{\,\mathrm{Ry}}
\def \pMpc	{\,\mathrm{pMpc}}
\def \cMpc	{\,\mathrm{cMpc}}

\def \ckpc	{\,\mathrm{ckpc}}
	
\def \Myr	{\,\mathrm{Myr}}

\def \Msun	{\,\mathrm{M_\odot}}
\def \um	{\,\mathrm{\mu{}m}}
\def \GqsoH	{\Gamma^\mathrm{HI}_\mathrm{QSO}}

\def \GuvbH	{\Gamma^\mathrm{HI}_\mathrm{UVB}}

\def \tQ	{t_\mathrm{Q}}
\def \tTO	{t_\mathrm{age}}
\def \tage	{t_\mathrm{age}}

\def \nlos	{n_\mathrm{los}}

\def \THETA	{\boldsymbol{\Theta}}

\newcommand{\lya}{{\rm Ly}\alpha}

\newcommand{\Nyx}{\textsc{Nyx} }
\newcommand{\Clamato}{\textsc{Clamato}{}}

\DeclareMathAlphabet{\mathsc}{OT1}{cmr}{m}{sc}
\DeclareRobustCommand{\ion}[2]{%
  \relax
  \ifmmode
    \ifx\testbx\f@series
      {\mathbf{#1\,\mathsc{#2}}}
    \else
      {\mathrm{#1\,\mathsc{#2}}}
    \fi
  \else
    \textup{#1\,{\mdseries\textsc{#2}}}%
  \fi
 }

\begin{document}

\lefthead{Mapping quasar light echoes in 3D with $\lya$ forest tomography}\righthead{T. M. Schmidt et al.}
\title{Mapping quasar light echoes in 3D with $\lya$ forest tomography}

\author{Tobias M. Schmidt\altaffilmark{1,2,7,*}, Joseph F. Hennawi\altaffilmark{1,2}, Khee-Gan Lee\altaffilmark{3,4}, Zarija Luki{\'c}\altaffilmark{4}, Jose O{\~n}orbe\altaffilmark{5}, Martin White\altaffilmark{4,6}}	
 
\altaffiltext{*}{e-mail: tschmidt@mpia.de}
\altaffiltext{1}{Department of Physics, University of California, Santa Barbara, CA 93106, USA}
\altaffiltext{2}{Max-Planck-Institut f\"ur Astronomie, K\"onigstuhl 17, D-69117 Heidelberg, Germany}
\altaffiltext{3}{Kavli Institute for the Physics and Mathematics of the Universe (IPMU), The University of Tokyo, Kashiwano-ha 5-1-5, Kashiwa-shi, Chiba, Japan}
\altaffiltext{4}{Lawrence Berkeley National Laboratory, CA 94720-8139, USA}
\altaffiltext{5}{Institute for Astronomy, Royal Observatory of Edinburgh, Blackford Hill, Edinburgh EH9 3HJ, United Kingdom}
\altaffiltext{6}{Department of Astronomy, University of California at Berkeley, B-20 Hearst Field Annex 3411, Berkeley, CA 94720, USA}
\altaffiltext{7}{Fellow of the International Max Planck Research School for Astronomy and
Cosmic Physics at the University of Heidelberg (IMPRS-HD).}

\begin{abstract}
The intense radiation emitted by luminous quasars dramatically alters the ionization state of their surrounding IGM. This so-called proximity effect extends out to tens of Mpc, and manifests as large coherent regions of enhanced Lyman-$\alpha$ (Ly$\alpha$) forest transmission in absorption spectra of background sightlines. Here we  present a novel method based on Ly$\alpha$ forest tomography, which is capable of mapping these quasar `light echoes' in three dimensions. Using a dense grid (10-100) of faint ($m_r\approx24.7\,\mathrm{mag}$) background galaxies as absorption probes, one can measure the ionization state of the IGM in the vicinity of a foreground quasar, yielding detailed information about the quasar's radiative history and emission geometry.  An end-to-end analysis -- combining cosmological hydrodynamical simulations post-processed with a quasar emission model, realistic estimates of galaxy number densities, and instrument + telescope throughput -- is conducted to explore the feasibility of detecting quasar light echoes.  We present a new fully Bayesian statistical method that allows one to reconstruct quasar light echoes from thousands of individual low S/N transmission measurements. Armed with this machinery, we undertake an exhaustive parameter study and show that light echoes can be convincingly detected for luminous ($M_{1450} < -27.5\,\mathrm{mag}$ corresponding to $m_{1450} < 18.4\,\mathrm{mag}$ at $z\simeq 3.6$) quasars at redshifts $3<z_\mathrm{QSO}<5$, and that a relative precision better than $20\,\%$ on the quasar age can be achieved for individual objects, for the expected range of ages between 1 Myr and 100 Myr. The observational requirements are relatively modest -- moderate resolution ($R\gtrsim750$) multi object spectroscopy at low $\rm{}S/N > 5$ is sufficient, requiring three hour integrations using existing instruments on 8m class telescopes.
\end{abstract}

\keywords{quasars: general -- intergalactic medium  -- reionization}

\section{Introduction}

Quasars are the the most-luminous non-transient sources in the Universe and can be observed throughout cosmic history \citep{Banados2016, Padovani2017}. They draw their immense power from accretion of matter onto a supermassive black hole (SMBH) and while every massive galaxy is expected to host a SMBH, most of them remain in a quiescent stage and only a small fraction appear as luminous quasars at any given time.  The processes which trigger quasar activity are presently unknown and models deliver a diversity of explanations for the sources of nuclear activity \citep[e.g][]{DiMatteo2005, Springel2005, Hopkins2007, Novak2011, Cisternas2011}.
Observationally, one would like to constrain the duration of bright quasar phases, the so called quasar lifetime $\tQ$. However, observations have so far not converged on a conclusive picture (see e.g. \citealt{Martini2004} for a review). For example, clustering measurements of quasars can constrain the integrated time galaxies (or cosmic halos) host quasars and suggest values between $10^7\yr$ and $10^9\yr$ \citep{Adelberger2005, Croom2005, Shen2009, White2012, Conroy2013}. Other techniques focusing on individual objects are more sensitive to the duration of the last quasar burst. For this, some authors claim rather short times between $10^4$ and $10^5\yr$ \citep{Schawinski2015, Eilers2017}, while others see indications for extended quasar activity up to a few times $10^7\yr$ \citep{Jakobsen2003, Goncalves2008, Worseck2006, Schmidt2017, Schmidt2018}.

Due to their high luminosity quasars have a profound impact on their environment on various scales. A better understanding of quasar activity cycles would therefore not only lead to a better understanding of the physics powering active galactic nuclei (AGN) and the growth of SMBH \citep{Soltan1982, Shankar2009, Kelly2010} but also shed light on the impact of AGN on their host galaxies, in particular quasar feedback which is usually invoked in galaxy formation simulations \citep[][] {DiMatteo2005,Hopkins2008ApJS175356H, Hopkins2008ApJS175390H}. In addition, quasars are the dominant source of hard ionizing photons ($E \gg 1\Ry$) in the Universe and it has been proposed that \textit{quasar flickering} could have a substantial impact on the ionization state of metals in the circumgalactic medium \citep{Oppenheimer2018, Segers2017}. On even larger scales, quasars dominate the metagalactic UV background at $z\lesssim3.5$
\citep{Faucher-Giguere2009, HaardtMadau2012, Khaire2018, Kulkarni2018}, which maintains the photoionization of the intergalactic medium (IGM) and drives the reionization of \ion{He}{ii}, where the latter directly couples the morphology of helium reionization to the emission properties of quasars \citep[e.g.][]{Davies2017,Schmidt2018}.

The impact of quasars on the IGM is probably also the most promising way to characterize their emission histories. Due to their large amount of ionizing photons, quasars create so called \textit{proximity  zones} in the IGM, megaparsec sized regions with enhanced photoionization and therefore reduced $\lya$ forest absorption. This \textit{proximity effect} can be observed in various ways.

The \textit{line-of-sight proximity effect} describes reduced IGM absorption in quasar spectra close to the quasar position \citep{Carswell1982, Bajtlik1988, Scott2000, DallAglio2008b, Calverley2011}.  Since this observation takes place \textit{along the light cone}, one can not directly probe the quasar luminosity at
past times. However, radiative transfer effects result in some sensitivity to the quasar emission history on scales comparable to the photoionization timescale. Because at $z<5$ the photoionization timescale is for \ion{H}{i} only $t_\mathrm{eq}\sim10^4\yr$, this method usually delivers only lower limits \citep{Eilers2017, Eilers2018}.  At higher redshift, during the epoch of reionization, $t_\mathrm{eq}$ is longer and a substantially neutral IGM can further delay the buildup of proximity zones, facilitating sensitivity to longer quasar ages \citep{Davies2018b,Davies2018c}. Alternatively, one can analyze the \ion{He}{ii} $\lya$ forest for which the photoionization rate is $\approx1000\times$ lower and therefore the photoionization time is longer by that amount \citep{Khrykin2016, Khrykin2018}.
 
Apart from the line-of-sight proximity effect, there is the \textit{transverse proximity effect} which comes into effect for close quasar pairs. Here, a background sightline passes close to a foreground quasar, probing the $\lya$ forest absorption in the vicinity of this foreground quasar.  The big advantage of this configuration is, that the pathlength from the foreground quasar to a point an the background sightline and from there to Earth is longer than the direct path from the foreground quasar to Earth. Therefore, the IGM absorption along the background sightline probes past times and is directly sensitive to the emission history of the foreground quasar. This enables constraints on the quasar age based on pure geometry and the finite speed of light \citep[see e.g.][]{Dobrzycki1991, Smette2002, Jakobsen2003, Adelberger2004, Hennawi2006, Hennawi2007, Furlanetto2011}. In addition, the background sightline probes the foreground quasars radiation from a different viewing angle than our vantage point from Earth. Quasar unification models predict that each quasar is obscured (e.g. by a dusty torus) toward a substantial part of the sky \citep{Antonucci1993, Urri1995, Netzer2015,Lusso2013}. The transverse proximity effect provides an opportunity to directly test these models and to constrain the emission geometry of individual quasars.

To date, the \ion{H}{i} transverse proximity effect has, for various reasons, thus far delivered inconclusive results \citep[e.g.][]{Liske2001, Schirber2004, Croft2004, Hennawi2006, Hennawi2007, Kirkman2008, Lau2016}. For example, at the redshift $z\approx3$ typical of most observations, the mean IGM transmission is already fairly high, and the enhancement resulting is relatively small, making it difficult to detect the proximity effect at this redshift. 
For \ion{He}{ii} however, the $\lya$ opacity at similar redshifts is much higher and a single foreground quasar can cause a dramatic increase in the \ion{He}{ii} $\lya$ forest transmission along a background sightline. The best example for this is the discovery of a strong \ion{He}{ii} proximity effect along the Q0302-003 sightline \citep{Heap2000, Jakobsen2003}, which enables joint constraints on the age and obscuration of this foreground quasar \citep{Schmidt2018}. 
However, \citet{Schmidt2018} analyzed other similar sightlines and found a diversity of quasar emission properties. In particular, the absence of transmission spikes for several foreground quasars suggests that these quasars are either very young or highly obscured. This degeneracy can not be broken as long as only single background sightlines are analyzed. Unfortunately, the number of \ion{He}{ii} sightlines is limited ($\approx$28) and the analysis of large quasar samples infeasible.

\begin{figure}
 \centering
 \includegraphics[width=\linewidth]{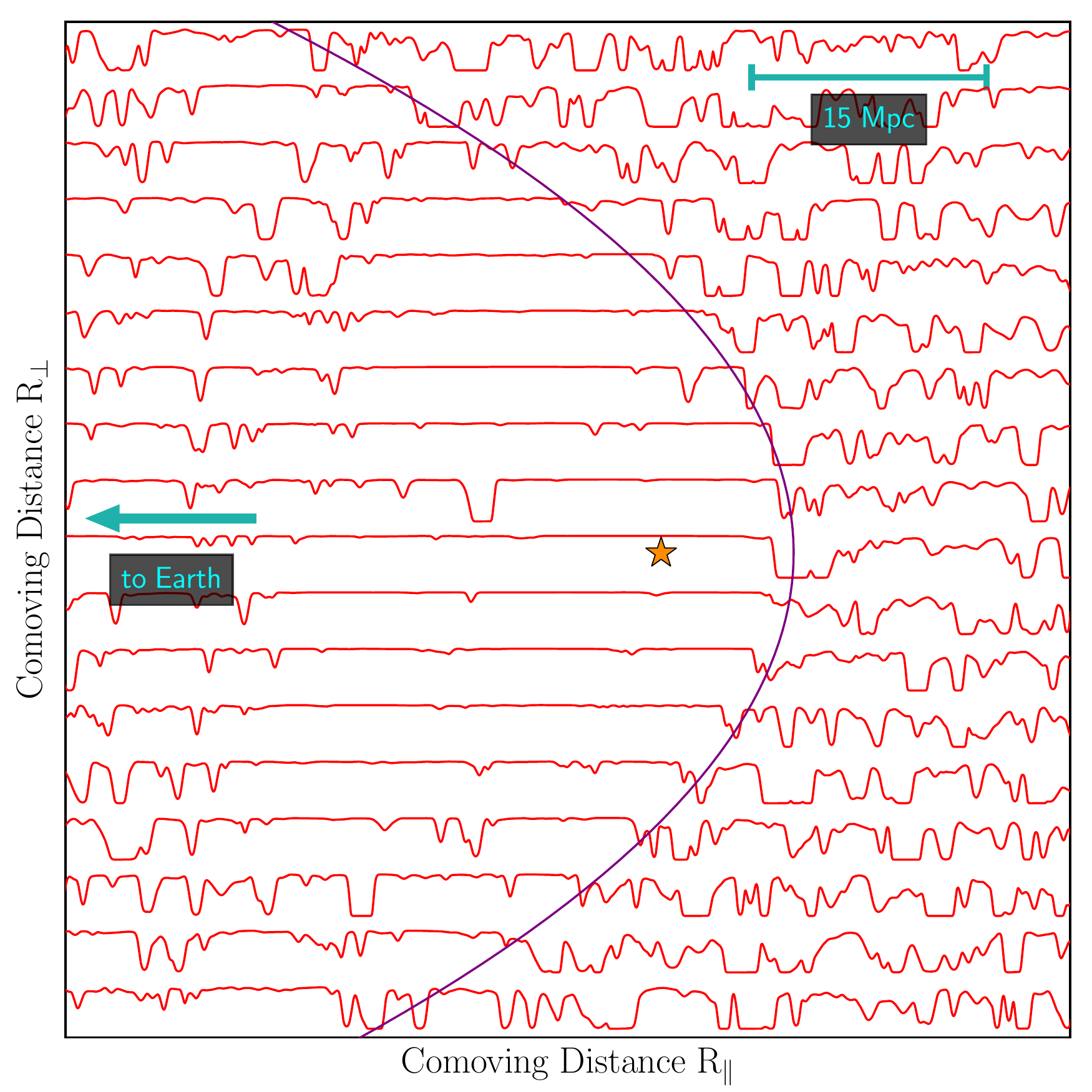}
 \caption{
 Illustration of the observational concept for mapping quasar light echoes with $\lya$ forest tomography. The UV radiation of a bright quasar (center) enhances the ionization state of the IGM in its surrounding. This proximity zone is then probed by many background sightlines (red). 
 The $\lya$ forest absorption along this ensemble of spectra clearly reveals the shape and structure of the proximity zone. In this case, the region influenced by the quasars radiation has a parabolic shape due to the finite age of the quasar (purple). See also \citet{Visbal2008}.
 This Figure is based on our models described in \S\ref{Sec:HydroSims}, but simplified and idealized to act as a sketch. A more realistic simulation of the proximity effect is shown in Figure~\ref{Fig:HI_TPE_Slice}.
 }
 \label{Fig:Sketch}
\end{figure}

A special variant of the transverse proximity effect does not rely on quasars as background source but uses faint (e.g. $r>24\Mag$) star forming galaxies. These are sufficiently abundant that the proximity zone of the foreground quasar can be probed via the $\lya$ forest absorption along many background sightlines.  This technique, first proposed by \cite{Adelberger2004} and \cite{Visbal2008}, allows one to map quasar light echoes in three dimensions and in much more detail
than possible with single background sightlines.  The concept is illustrated in Figure~\ref{Fig:Sketch} which clearly shows how the parabolic-shaped appearance of the quasar proximity zone, which is caused by the finite age of the quasar (see \S~\ref{Sec:QuasarLifetime}), can be seen in the $\lya$ forest absorption along the background sightlines. Despite the great potential, this tomographic mapping of quasar light echoes has so far never been attempted in practice, probably because the observational requirement were judged to be too challenging.

Tomographic reconstructions of the large-scale structure of the Universe using the Ly$\alpha$ forest absorption in the spectra of faint background galaxies were recently pioneered by \citet{Lee2014a, Lee2014b}. They showed that  at $z\sim2.4$, using faint $r>24\Mag$ star forming galaxies as background sources delivers sightline densities around $700\,\mathrm{deg}^{-2}$, which is sufficient to interpolate between sightlines and to reconstruct a tomographic map of the IGM absorption on Mpc scales. By clearly assessing the observational requirement, \cite{Lee2014a} determined that $\lya$ tomography is indeed possible with current-generation facilities, in particular 8--10\,m class telescopes and existing multi-object spectrographs. Since then, the  COSMOS Lyman-Alpha Mapping And Tomography Observations (\Clamato{}) survey has proven the feasibility in practice and delivered the first tomographic map of the IGM \cite{Lee2014b, Lee2017}. Prime objective of the \Clamato{} survey is to map the large-scale structure of the Universe to find e.g. protocluster \cite{Stark2015a, Lee2016} and to study the cosmic web \citep{Stark2015b, Krolewski2018}. Similar techniques, based however on SDSS/BOSS spectra, were employed by \citet{Cai2015, Cai2017} to map large-scale overdensities. 

In light of these developments, this paper revisits the question of mapping quasar light echoes with Ly$\alpha$ forest tomography. Our aim is to demonstrate feasibility, assess sensitivity, and determine the optimal observing strategy by conducting an end-to-end analysis of the experiment, starting from the observational requirements, computation of realistic models, and finally fully Bayesian inference of parameter constraints. To keep the complexity of this pilot study at an acceptable level and to limit the computational expense, we adopt a simple isotropic \textit{lightbulb} model for the quasar emission and focus only on measurements of the quasar age. In the future, we will relax these assumptions and consider more realistic anisotropic emission from quasars as well as more complex lightcurves.

The structure of this paper is as follows. We summarize all relevant observational parameters like quasar luminosity, instrument properties (sensitivity, spectral resolution, field-of-view), and achievable background sightline density in \S\ref{Sec:ObservationalSetup}.
We will then describe our models of the 3D proximity effect, which are based on state-of-the-art cosmological hydrodynamical simulations postprocessed with a quasar emission model  (\S\ref{Sec:HydroSims}, \S\ref{Sec:SimulatedData}). To compare observational data to our models and infer posterior probability distributions for the quasar age, we develop a sophisticated Bayesian method based on likelihood-free inference (\S\ref{Sec:LikelihoodComputation}). Finally, we apply this analysis pipeline to mock observations and  determine the achievable precision on quasar age and assess dependencies on various observational parameters (\S\ref{Sec:Results}), allowing us to choose the optimal observing strategy.

Throughout the paper we use a flat $\Lambda$CDM cosmology with $H\mathrm{_0 =68.5\,km \: s^{-1} \: Mpc^{-1}}$, $\mathrm{\Omega_\Lambda = 0.7}$, $\mathrm{\Omega_m = 0.3}$ and $\mathrm{\Omega_b = 0.047}$ which has been used for the computation of the \Nyx hydro simulation and is broadly consistent with the \citet{Planck2018VI} results. We use comoving distances and denote the corresponding units as $\cMpc$. In this paper, we consider a simple lightbulb model for the quasar lightcurve in which the quasar turns on, shines with constant luminosity for its full lifetime $\tQ$ until it turns off. This timespan is however different from the quasar age $\tTO$, which describes the time from turning on until emission of the photons that are observed on Earth today (see Figure~\ref{Fig:LightBulb}). Magnitudes are given in the AB system \citep{Oke1983}.

\section{Observational Setup}
\label{Sec:ObservationalSetup}

To set the stage for our undertaking and to define the observational framework of our study, we discuss all relevant observational parameters in this section. This includes the luminosity of potential quasar targets, achievable background sightline densities, required spectral resolution, exposure times, field-of-view (FoV), and other elements. For most of these parameters we will motivate some initial estimates and build up a fiducial observing strategy.  We will later in \S~\ref{Sec:Results} explore in detail how the quality of the inferred quasar properties depends on these choices and show how the strategy can be optimized.

\label{Sec:OptimalQuasarRedshift}

A key question for mapping quasar light echoes with $\lya$ forest tomography is the optimal quasar redshift to operate at.  At low redshift, the average IGM transmission is quite high, e.g. 85\% at $z=2.2$, and in consequence even the brightest foreground quasar can only cause a very limited enhancement in the $\lya$ forest transmission. In such conditions, the stochastic nature of the $\lya$ forest absorption and unavoidable uncertainties in estimating the continuum of the background sources might conceal the transmission enhancement.
With increasing redshift and thus lower mean IGM transmission (e.g. 45\% at $z=3.8$), foreground quasars can cause a stronger transmission
enhancement that is easier to detect in a tomographic map. This clearly motivates working at higher redshifts.  
On the other hand, the number density and brightness of available background galaxies drops
dramatically with increasing redshift. One therefore has to accept a much sparser sampling of the proximity region by background
sightlines, work with lower $\rm{}S/N$ spectra or substantially increase the exposure times. 
Finding the best compromise between strong transmission enhancement at high redshift and fine sampling of the tomographic map at low redshift is one of the primary intentions of this work and requires a detailed assessment of the relevant observational parameters.

\begin{deluxetable}{lcrc}
\tablecolumns{4}
\tablewidth{0.95\linewidth}
\tablecaption{
Key Properties of Spectrographs Usable for this Project. 
}
\tablehead{
\colhead{Instrument}	& \colhead{FoV} & \colhead{$R$\;\tablenotemark{a}} & \colhead{$r_\mathrm{lim}$\;\tablenotemark{b}}
}
\startdata
VLT/FORSII	& $6.8^\prime \times 6.8^\prime$	& 945 	& $24.7\,\Mag$	\\
Keck/LRIS	& $7.8^\prime \times 6.0^\prime$	& 1435 	& $24.4\,\Mag$	\\
Keck/DEIMOS	& $16.7^\prime \times 5.0^\prime$	& 1852	& $24.5\,\Mag$	\\
Subaru/PFS	& $78^\prime$ diameter			& 2300	& t.b.d
\enddata
\tablenotetext{a}{Resolving Power at $5600\,\mathrm{\AA}$ and with $1.0"$ slit}
\tablenotetext{b}{Limiting $r$~band magnitude to reach $\rm{}S/N_{1000}=5.0$ at $5600\,\mathrm{\AA}$ in $t_\mathrm{exp}=10\,\mathrm{ks}$ }
\label{Tab:Spectrographs}
\end{deluxetable}

\subsection{Quasar Luminosities}

\begin{figure}
 \centering
 \includegraphics[width=\linewidth]{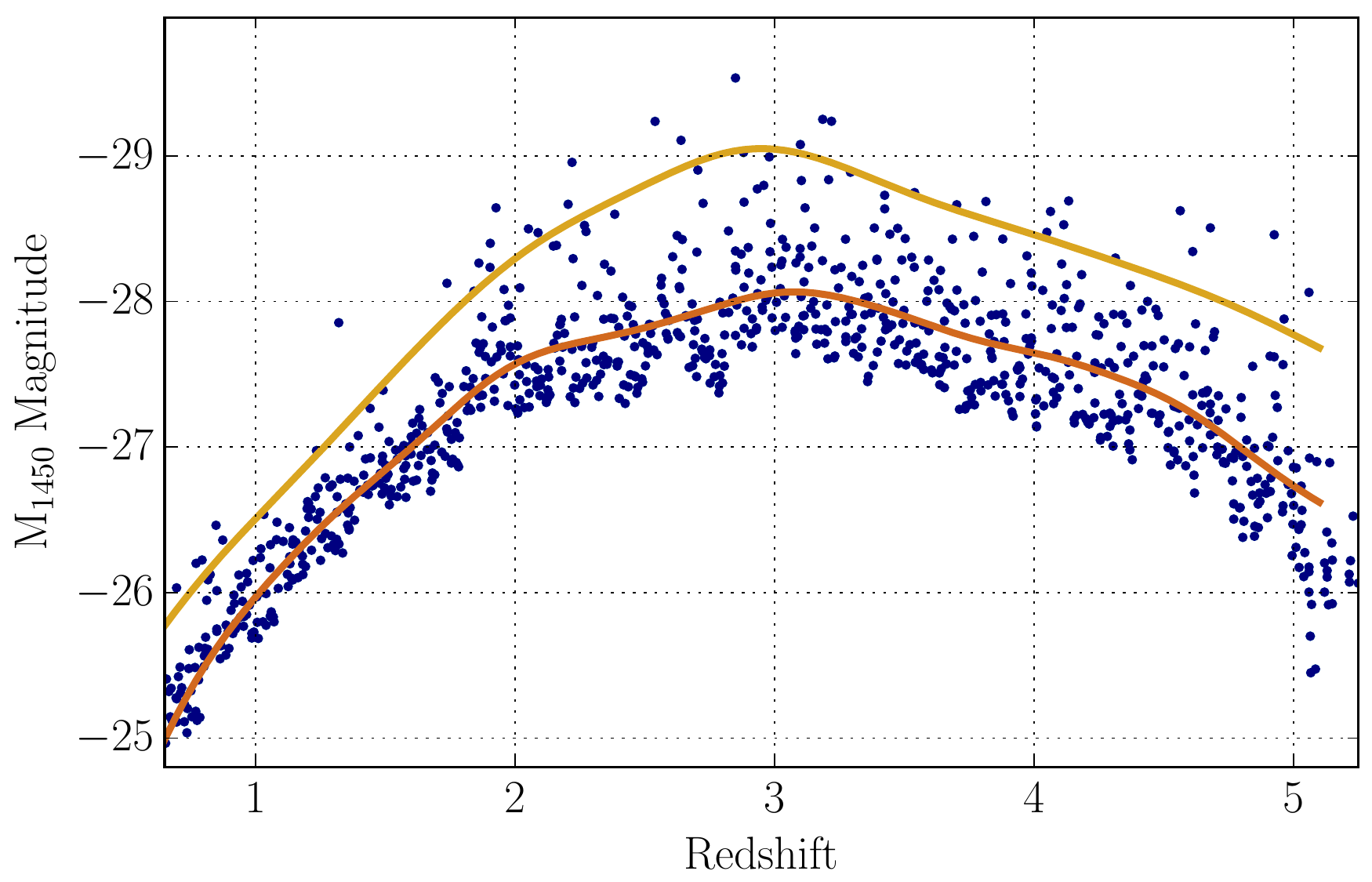}
 \caption{Most luminous quasars from the SDSS/BOSS DR14Q quasar catalog \citep{Paris2017}. For each $\Delta{}z = 0.05$ bin we have selected the ten brightest quasars. Solid lines show running averages of the full selected sample (dark line) and just the most luminous quasars in each redshift bin (light line).
 }
 \label{Fig:BrightestQuasars}
\end{figure}

One would ideally target the brightest quasars at a given redshift since these cause the strongest enhancement in transmission and have the largest proximity zone.  We have therefore compiled a collection of the most luminous quasars at each redshift based on the SDSS DR14Q spectroscopic quasar catalog \citep{Paris2017}. For each redshift bin of size $\Delta{}z=0.05$, we selected the ten brightest objects and show the resulting sample in Figure~\ref{Fig:BrightestQuasars}. Conversion from observed SDSS $i$-band magnitude to monochromatic luminosity at $1450\,\mathrm{\AA}$ ($M_{1450}$) is based on the \citet{Lusso2015} quasar template and Galactic extinctions from \citet{Schlafly2011}\footnote{https://irsa.ipac.caltech.edu/applications/DUST/}.

Clearly, the most luminous quasars exist around $z \approx 3$ and reach absolute UV magnitudes around $M_{1450}=-29\Mag$. For higher redshifts, the peak luminosity slowly decreases to $M_{1450}=-28\Mag$ at $z=5$ while it steeply drops for redshifts below $z<2.5$.  
By computing a running average (Gaussian filter of $\Delta{}z=0.2$ width) of the brightest quasar per $\Delta{}z=0.05$ redshift bin, we obtain a suitable representation of the evolution of the most luminous quasars in the universe. These are indeed the ideal targets for our experiment and for the rest of this paper we will use this smooth function (shown in Figure~\ref{Fig:BrightestQuasars} in light brown color) as the fiducial quasar luminosity.

We stress that other quasar catalogs \citep[e.g][]{Schmidt1983, VeronCetty2010, Flesch2015, Schindler2018}  list  additional ultraluminous quasars. However, including these affects predominantly redshifts $z_\mathrm{QSO}<2$ and does not change the overall picture at the redshifts $2 < z_\mathrm{QSO} < 5$ for which the $\lya$ forest is accessible from the ground.

\subsection{Field-of-View} 

\begin{figure}
 \centering
 \includegraphics[width=\linewidth]{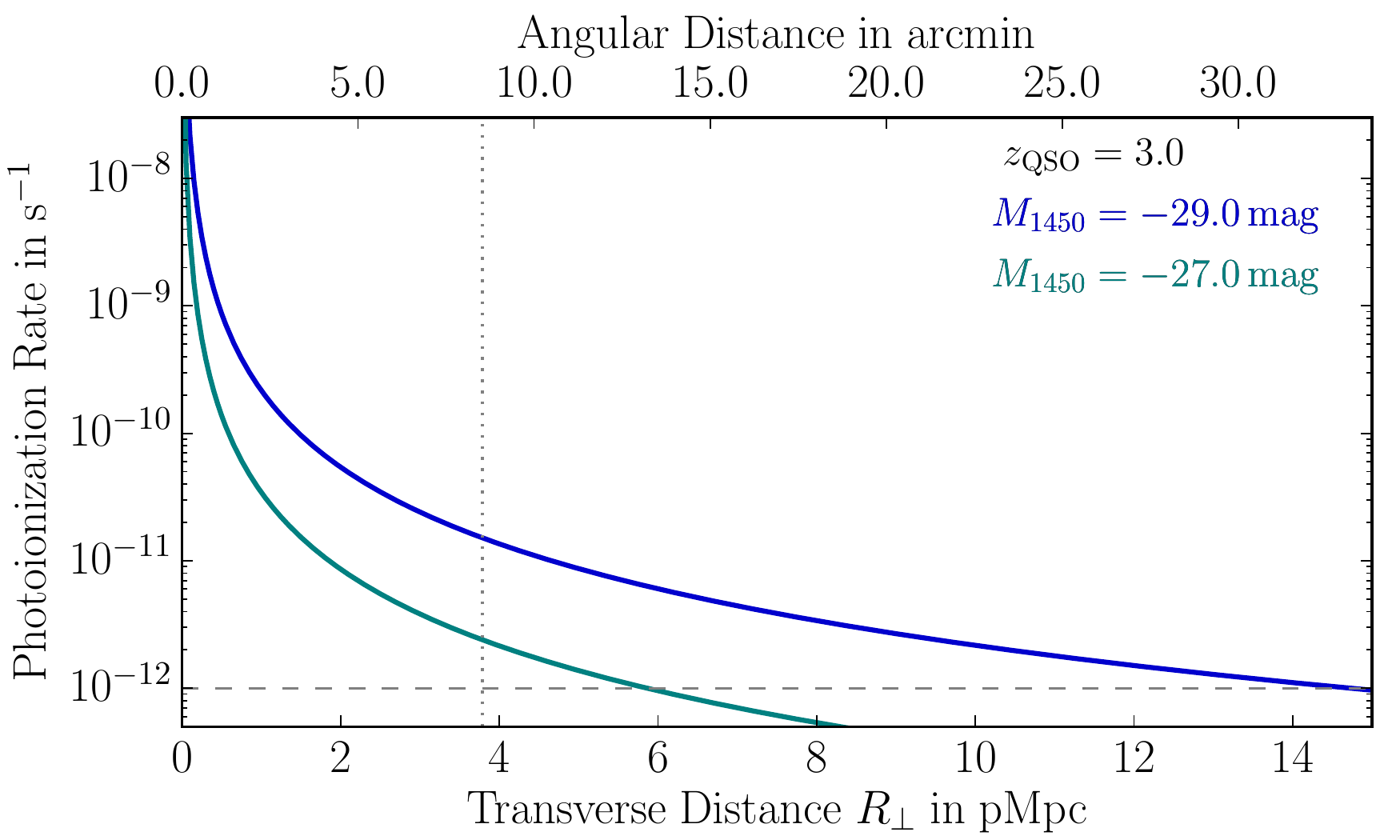}
 \caption{Expected quasar photoionization rate (see \S~\ref{Sec:Quasar_Ionization_Model} for the derivation) for two $z=3.0$ quasars of different luminosity  as function of transverse distance $R_\perp$, compared to the metagalactic UV background of $\GuvbH\approx10^{-12}\,\mathrm{s}^{-1}$ \citep{Becker2013b}. 
 We show angular and proper distance (instead of comoving) since in these coordinates the size of the proximity zone has only a weak dependence on redshift. The vertical dotted line indicates the radius of the FoV adopted throughout this paper.
 }
 \label{Fig:GammaQSO}
\end{figure}

The region of interest around the foreground quasar is clearly set by the size of its proximity region, i.e. the region where the photoionization rate of the quasar $\GqsoH$ substantially exceeds the UV background photoionization rate $\GuvbH$. 
The optical depth in the $\lya$ forest scales approximately inversely proportional to the ionization rate. Therefore, naively
a $\sim 100\%$ increase in $\Gamma_\mathrm{total}^\mathrm{HI}$ due to the quasars ionizing flux should result in a detectable
effect. 

In Figure~\ref{Fig:GammaQSO}, we show the expected quasar photoionization rate as a function of transverse distance $R_\perp$, assuming a fiducial quasar luminosity of $M_{1450}=-29\Mag$, consistent with the most luminous quasars in the universe (see Figure~\ref{Fig:BrightestQuasars}), the \citet{Lusso2015} quasar template and no Lyman limit absorption by the IGM (see Equation~\ref{Eq:Fnu_QSO} and \ref{Eq:Q_QSO} in \S\ref{Sec:Quasar_Ionization_Model} for more details). The UV background photoionization rate is of order $10^{-12}\,\mathrm{s}^{-1}$ \citep{Becker2013b}. Therefore, hyperluminous quasars dominate the photoionization rate out to $\approx14\pMpc$ distance, corresponding to $30^\prime$ or $56\cMpc$ at $z_\mathrm{QSO}=3$ (Figure~\ref{Fig:GammaQSO}). 
However, as listed in Table~\ref{Tab:Spectrographs}, the field-of-view (FoV) of classical multi-object spectrographs is usually  $\ll10^\prime$. This implies that, for typical spectrographs, covering the full extent of a proximity zone would require multiple pointings. But to remain efficient, one might rather focus on the central region where the quasar radiation will cause the strongest impact on the $\lya$ IGM transmission. The exception is the Subaru Prime Focus Spectrograph (first light in 2021) which will have a circular field-of-view with $1.3\,\mathrm{degree}$ diameter that could cover the full proximity region with a single pointing.
However, our reference concept focuses on the capabilities of currently existing instruments and therefore assumes a circular FoV with $16^\prime$ diameter. This could be covered by a $2\times2$ VLT/FORS\,II or Keck/LRIS mosaic, or a $3\times1$ mosaic with DEIMOS. Within this region the photoionization rate of an $M_{1450}=-29\Mag$ quasar exceeds the UV background by more than an order of magnitude and will strong alter the $\lya$ IGM transmission. The usefulness of larger fields will be explored later in \S\ref{Sec:ResFoV}.  

For the calculation underlying Figure~\ref{Fig:GammaQSO} we have assumed $z_\mathrm{QSO}=3$, but we stress that the amplitude of the UV background as well as the conversion from proper transverse distance to angular size depends only weakly on redshift. Therefore, Figure~\ref{Fig:GammaQSO} is representative for the full redshift range $2 < z_\mathrm{QSO} < 5$ we consider in this paper.

\subsection{Spectral Resolution}
\label{Sec:R_Requirement}

As already pointed out by \citet{Adelberger2004}, peculiar velocities pose a substantial limitation to tomographic quasar light echo measurements, since they introduce non-trivial distortions of the order of a few $100\,\mathrm{km\,s^{-1}}$ into the reconstructed map. Without a priori knowledge of the density field, which sources these motions,  it is of little benefit to take spectra with substantially better resolution than this velocity scale. Velocities of $300\,\mathrm{km\,s^{-1}}$ correspond to a resolving power of $R = \frac{\lambda}{\Delta{}\lambda} = 1000$ or  $3.9\cMpc$ at $z=3$. For now, we take this as the reference resolution and later explore in  detail, how the fidelity of our reconstructed tomographic map depends on the spectral resolution of the initial spectra (\S~\ref{Sec:ResResolvingPower}).

\subsection{Required $\rm{}S/N$ and Exposure Times}
\label{Sec:SN_Requirement}

The stochastic nature of the $\lya$ forest absorption causes substantial fluctuations when measuring the mean IGM transmission. For
example, an $\approx4\cMpc$ long chunk of a spectrum the scatter about the mean transmission is $\approx20\%$ (see Figure~\ref{Fig:HI_FluxStatistic}).
There is therefore limited gain in obtaining high $\rm{}S/N$ spectra, since for $\rm{}S/N\gtrsim5$, intrinsic fluctuations in the IGM absorption dominate the measurement uncertainty. The exact numbers depend of course on the IGM mean transmission and therefore redshift, however, the in general quite modest requirements on spectral is one of the key factors that makes $\lya$ forest tomography feasible with current generation telescopes \citep{Lee2014a}.

To allow for easier comparison between different instruments and spectral resolutions, we define $\rm{}S/N_{1000}$ as the achieved $\rm{}S/N$ per $R=1000$ or $300\,\mathrm{km\,s^{-1}}$ resolution element, independent of the actual resolution or pixel scale of the observations.%
\footnote{ Spectrographs might have a higher spectral resolution (see Table~\ref{Tab:Spectrographs}) and certainly a finer pixel scale to sample the line-spread function with up to 8 pixels. Therefore, a $300\,\mathrm{km\,s^{-1}}$ wide chunk of a spectrum will be sampled by several ($N$) pixels and the actual $\rm{}S/N$ per pixel will be lower by $\sqrt{N}$. }
Specifying $\rm{}S/N_{1000}$ ensures that for higher resolving power the light is more dispersed and finer sampled but the overall number of detected photons and therefore the required exposure time to reach a certain $\rm{}S/N_{1000}$ is conserved. The choice of $R=1000$ as reference resolution is arbitrary, however it is close to the minimum resolving power we require for this project. 
In addition, to be independent of the actual IGM absorption, we define $\rm{}S/N_{1000}$ as the continuum-to-noise in the region of the $\lya$ forest .
  
\begin{figure}
 \centering
 \includegraphics[width=\linewidth]{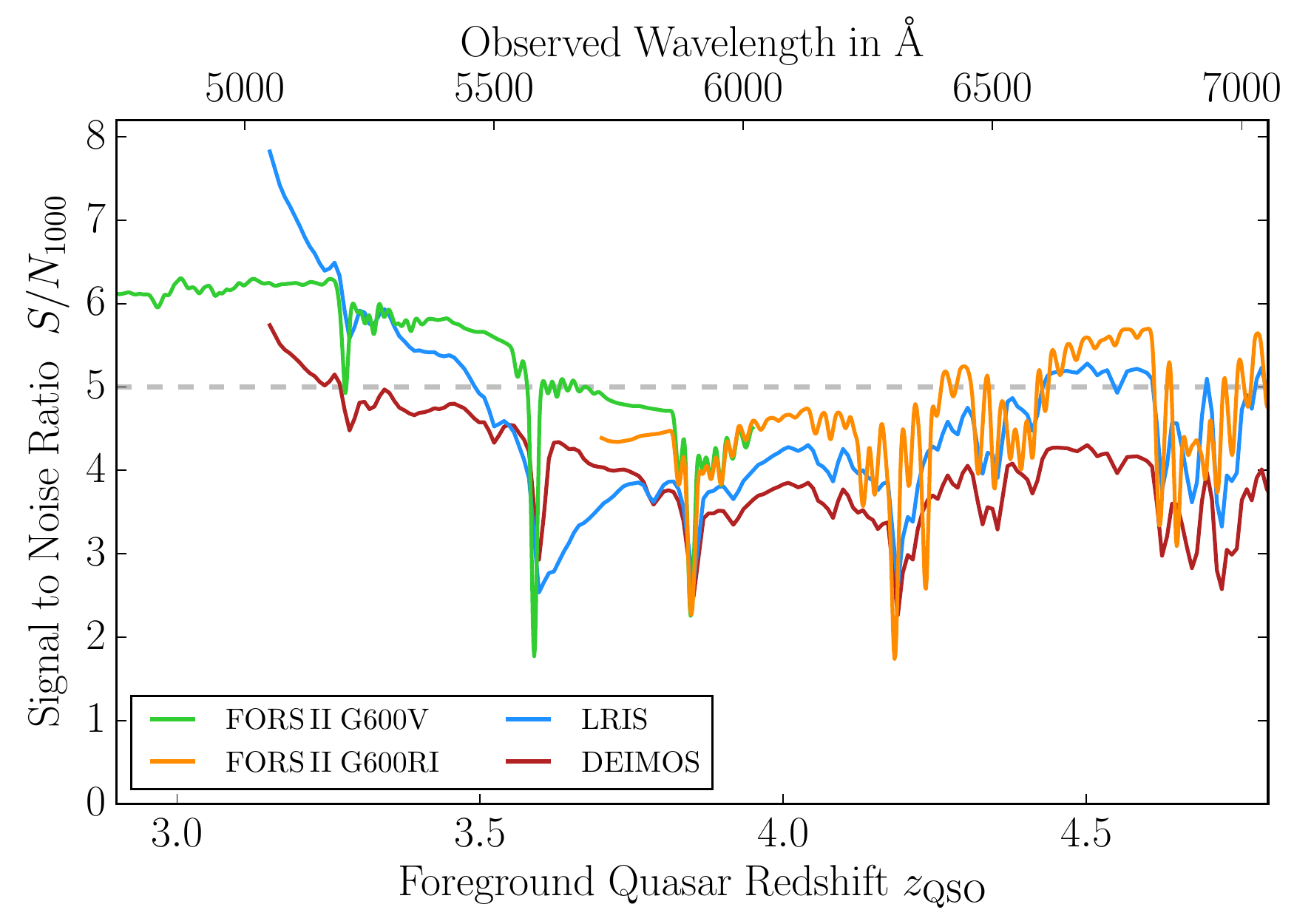}
 \caption{Sensitivity of different spectrographs as function of wavelength. Shown is the expected $S/N_{1000}$ at various foreground quasar redshifts achieved for an $r=24.7\Mag$ background galaxy in a $t_\mathrm{exp}=10\,\mathrm{ks}$ exposure. 
 For FORS\,II, we show the $S/N_{1000}$ calculation for a blue setup (grating G600V, E2V CCD) and a red setup (grating G600RI, MIT CCD). All calculations assume no Moon contribution.
 For LRIS we show the 400/4000 + 600/7500 and for DEIMOS the 900ZD setup 
 }
 \label{Fig:ETC_Instruments}
\end{figure}

In Table~\ref{Tab:Spectrographs} we have listed the approximate limiting $r$-band magnitude to reach a $\rm{}S/N_{1000}=5$ within a $10\,\mathrm{ks}$ exposure for different spectrographs. This calculation assumes the background galaxies have a power-law spectrum of the form $f_\lambda \propto \lambda^{-1.4}$ \citep{Bouwens2009}.
The limiting magnitude $r_{\rm lim}$ quoted is that which yields a continuum $\rm{}S/N_{1000}=5.0$ in the $\lya$ forest at $5600\,\mathrm{\AA}$, corresponding to a quasar redshifts of $z_\mathrm{QSO}\approx3.6$.  We have chosen to parametrize the apparent magnitude of the background galaxies in the $r$-band filter since it is conveniently observable and for $z_\mathrm{bg}<3.8$ samples the UV continuum of the galaxies redwards of $\lya$. For higher redshifts, one technically has to specify $i$-band magnitudes to avoid contamination by the $\lya$ forest. However, for the purpose of this work this is of no concern since we anyway specify unabsorbed continuum magnitudes.

The sensitivity of spectrographs is in general wavelength dependent and the $\rm{}S/N$ at the quasar position achieved in a fixed exposure time (or vice versa the limiting magnitude) depends -- even for the identical background galaxy -- on the redshift of the foreground quasar. We show this dependence for FORS\,II, LRIS and DEIMOS in Figure~\ref{Fig:ETC_Instruments}. The exposure time estimates are based on the Keck\footnote{\url{http://etc.ucolick.org/web_s2n/lris}, \url{ http://etc.ucolick.org/web_s2n/deimos}} and ESO\footnote{\url{https://www.eso.org/observing/etc/bin/gen/form?INS.NAME=FORS+INS.MODE=spectro}, Version P102.5} exposure time calculators and assume good but realistic conditions%
\footnote{FORS\,II: E2V blue detector and G600V grating or MIT red detector and G600RI grating, Airmass=1.2, Fractional Lunar Illumination (FLI) =0.0, Seeing=0.7" (47\% chance), $\mathrm{Slit=1"}$, $f_\lambda \propto \lambda^{-1.4}$}.
However, the dependence of achieved $\rm{}S/N_{1000}$ on wavelength is relatively weak (see Figure~\ref{Fig:ETC_Instruments}) and can to some degree mitigated by using either red or blue optimized instrument setups.
To not complicate matters any further, we ignore this wavelength dependence for the rest of our study and simply assume that for a limiting magnitude of $r_\mathrm{lim}=24.7$ a $\rm{}S/N_{1000}=5.0$ can be achieved in $t_\mathrm{exp} \approx 10\,\mathrm{ks}$, independent of quasar redshift. 
In practice, when planning actual observations, the true sensitivity of the instruments has to be taken into account and the exposure times, limiting magnitudes or $\rm{}S/N$ ratios adjusted accordingly. If adjustments to the observational parameters are necessary, these scale approximately like $ t_\mathrm{exp} \propto \left( \, \rm{}S/N_{1000} \, \right)^{2}$ and ${\rm{}S/N_{1000}} \propto 10^{\,-\frac{2}{5} \: m_\mathrm{r}}$ as long as the objects are substantially fainter than the sky brightness.

\subsection{Background Sightline Density}
\label{Sec:SightlineDensity}

\begin{figure}
 \centering
 \includegraphics[width=\linewidth]{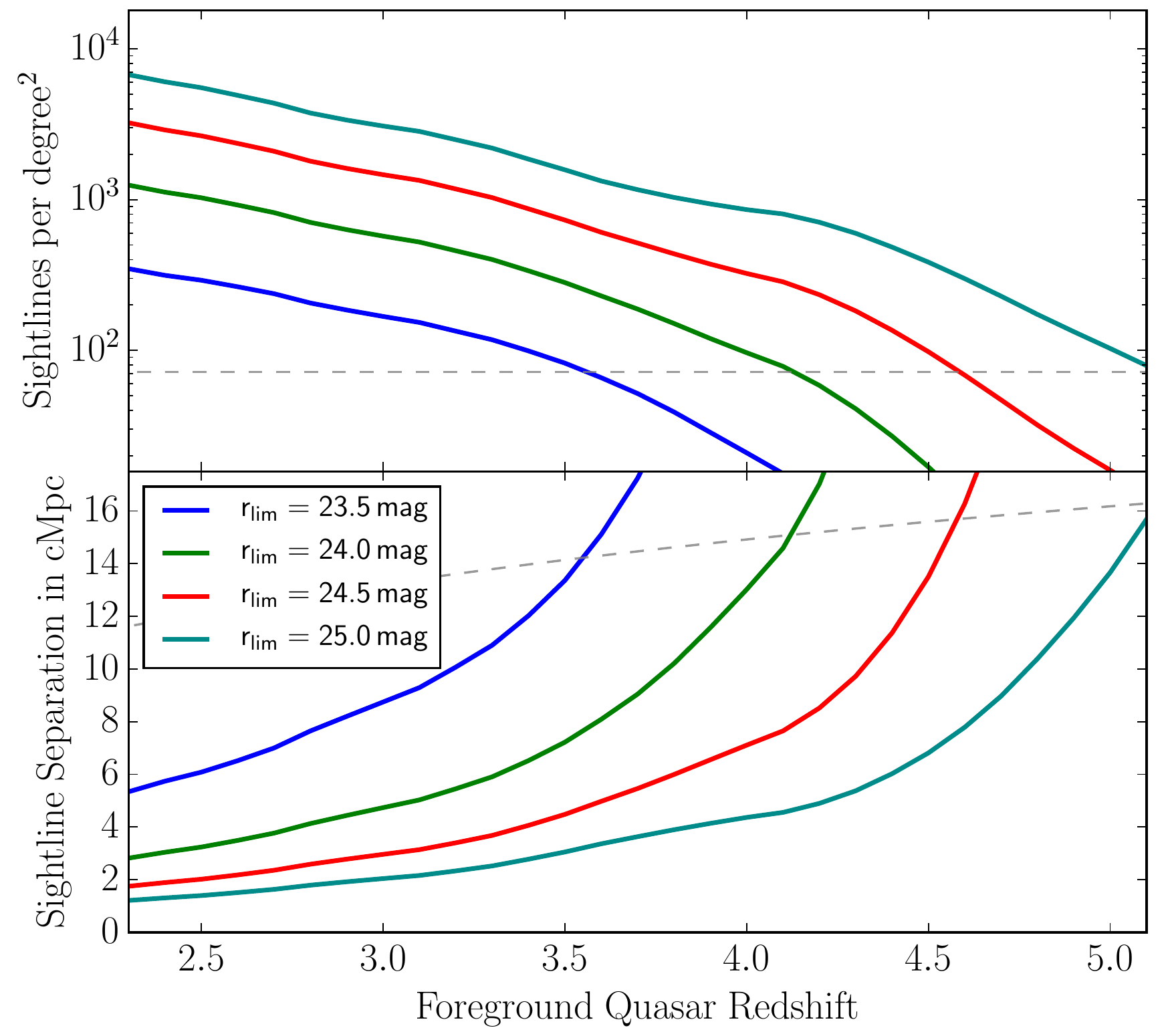}
 \caption{Achievable sightline density for different limiting SDSS~$r$ band magnitudes. Based on the \citet{Cucciati2012} and \citet{Bouwens2015} luminosity functions and assuming a power-law spectra of $f_\lambda \propto \lambda^{-1.4}$ \citep{Bouwens2009} for the background galaxies. 
 The shown limiting magnitudes correspond to exposure times of approximately $1000$, $2700$, $7000$ and $17000\,\mathrm{s}$.
 The dashed lines indicate the limit of one sightline per $50\,\mathrm{arcmin}^2$, approximately the LRIS or FORS\,II FoV.
 }
 \label{Fig:SightlineDensity}
\end{figure}

The most crucial factor for our experiment is probably the achievable density of background sightlines $\nlos$. We estimate this closely following the approach outlined in \citet{Lee2014a}. Based on a luminosity function $\Phi(z_{bg}, m)$, which specifies the number of galaxies per luminosity and comoving volume, the sightline density is given by
\begin{equation}
 \nlos = \int_{z_1}^{z_2} \int^{m_\mathrm{lim}}_{-\infty} \Phi(z_\mathrm{bg}, m) \; dm \frac{dl_\mathrm{c}}{dz_\mathrm{bg}} \; dz_\mathrm{bg}
 \label{Eq:nlos},
\end{equation}
where $m_\mathrm{lim}$ is the limiting apparent magnitude of our survey and $dl_\mathrm{c}$ the comoving line element along the line-of-sight. Background galaxies can contribute to the tomographic map at the quasar redshift $z_\mathrm{QSO}$ if their redshift $z_\mathrm{bg}$ falls in the redshift interval $z_1<z_\mathrm{bg}<z_2$ defined by 
\begin{equation}
 (1+z_\mathrm{QSO}) \; \lambda_\mathrm{Ly\alpha} = ( 1 + z_i) \; \lambda_i
\end{equation}
in which $\lambda_1$ and $\lambda_2$ denote -- in rest wavelengths -- the usable part of the background spectra, i.e. $\lambda_1 \approx \lambda_\mathrm{Ly\alpha} = 1216\,\mathrm{\AA}$ and $\lambda_2 \approx \lambda_\mathrm{Ly\beta} = 1025\,\mathrm{\AA}$.

We use the luminosity functions of \citet{Cucciati2012} for $z\leq4$ and combine it with the \citet{Bouwens2015} measurements at higher redshifts. In both cases, we use the analytic Schechter representation of the luminosity function and interpolate the function parameter between redshifts. 
To convert from the absolute UV magnitude specified around $1600\,\mathrm{\AA}$ in the luminosity functions to the apparent magnitude in our observed bandpass (SDSS $r$ band), we use the standard conversion \citep{Hogg1999} and assume a galaxy SED of the form $f_\lambda \propto \lambda^{-1.4}$ \citep{Bouwens2009}.

We show the result in Figure~\ref{Fig:SightlineDensity}, expressed once in terms of  sightlines per square degree and once as average comoving separation between sightlines. Clearly, fainter limiting magnitudes allow a higher sightline density and a finer sampling of the tomographic map. However, the achievable density of background sources drops rapidly with increasing quasar redshift. At $z_\mathrm{QSO}=2.5$ an average sightline separation of about $3\cMpc$ can be reached when only considering $r < 24\Mag$ background galaxies. At $z_\mathrm{QSO}=3.5$, one has to go half a magnitude deeper and still only reaches an average sightline separation of $4.5\cMpc$. 
For $z_\mathrm{QSO}=4.5$, even with background galaxies as faint as $r < 25\Mag$, average separations will be larger than $6\cMpc$.

\subsection{Summary of Observational Parameters}

In the sections above we have collected all dependencies of the observational parameters relevant for our $\lya$ forest tomography project. This now allows us to explore the parameter space by varying
single quantities like the desired $\rm{}S/N_{1000}$, the field-of-view observed, or the spectral resolution. This also also allows us to explort certain paths through the parameter space, e.g. vary the foreground quasar redshift $z_\mathrm{QSO}$ while simultaneously adjusting to the correct quasar brightness, limiting magnitude and therefore background sightline density, to keep the required exposure time constant. We will use this later in \S\ref{Sec:Results} find the optimal observing strategy for the project.

\section{Models / Simulations}
\label{Sec:HydroSims}

Given the observational framework outlined above, we create realistic models of the $\lya$ forest in the vicinity of bright quasars. Our models are based on outputs of cosmological hydrodynamical simulations which we postprocess with a photoionization model that explicitly incorporates finite quasar ages. From these we create mock $\lya$ forest spectra that resemble a given (e.g. the actually observed) pattern of background sightlines in the vicinity of a foreground quasar. In a final step, we forward model observational effects, in particular finite spectral resolution, finite $\rm{}S/N$, and continuum fitting errors.  The overall scheme is nearly identical to the one we developed in \citet{Schmidt2018} to create \ion{He}{ii} $\lya$ forest spectra.

\subsection{\Nyx Cosmological Hydrodynamical Simulations}

We use simulations computed with the Eulerian hydrodynamical simulation code \Nyx \citep{Almgren2013, Lukic2015}.  The simulation box has a large size of $100\,h^{-1}\cMpc$, required to capture the full extent of the proximity zone of hyperluminous quasars. The hydrodynamics is computed on a fixed grid of $4096^3$ resolution elements and the same number of dark matter particles is used to compute the gravitational forces. This results in a resolution of $36\ckpc$ per pixel, sufficient to resolve the \ion{H}{i} $\lya$~forest at $2.0<z<5.0$ \citep{Lukic2015}.  The simulation runs make no use of adaptive mesh refinement since the \ion{H}{i}~$\lya$~forest signal originates from the majority of the volume \citep{Lukic2015}. Refining the resolution in the dense regions at the expense of underdense regions is therefore not beneficial for our case.  Also, since the prime objective of the simulation is IGM science, no star or galaxy formation prescriptions were included.  The simulations were run using a homogeneous, optically thin UV background with photoionization and heating rates from \citet{HaardtMadau2012}. As described below, we rescale the \ion{H}{i} photoionization rates to closely match the observed mean transmission but keep the thermal structure unchanged.  We have simulation outputs available at redshifts $z=2$, 2.5, 3.0, 3.5, 4.0, 4.5 and 5. Depending on the desired foreground quasar redshift, we take the snapshot closest in redshift and extract density, velocity and temperature along skewers. These will later be post-processed to simulate the observed \ion{H}{i} Ly$\alpha$ transmission along background sightlines.

For the host dark matter halos of the foreground quasars we choose halos with masses $\gtrsim 10^{12}\Msun$ halos, which is the minimum mass of quasar host halos as indicated from clustering studies\citep[e.g][]{Richardson2012, White2012}. As described in more detail in \citet{Sorini2018}, halos in the \Nyx simulations are identified by finding topologically connected components above
138$\times$ mean density (Luki{\'c} et al. in prep.). This gives similar results to the particle-based friends-of-friends algorithm \citep{Davis1985}. From the \Nyx halo catalog we select all halos
with mass larger $> 10^{12}\Msun$ ($>7\times10^{11}\Msun$ for $z_\mathrm{QSO} \geq 4.8$) and randomly draw halos from this subset.

\subsubsection{Sightline Pattern and Skewer Extraction}

\begin{figure}
 \centering
 \includegraphics[width=\linewidth]{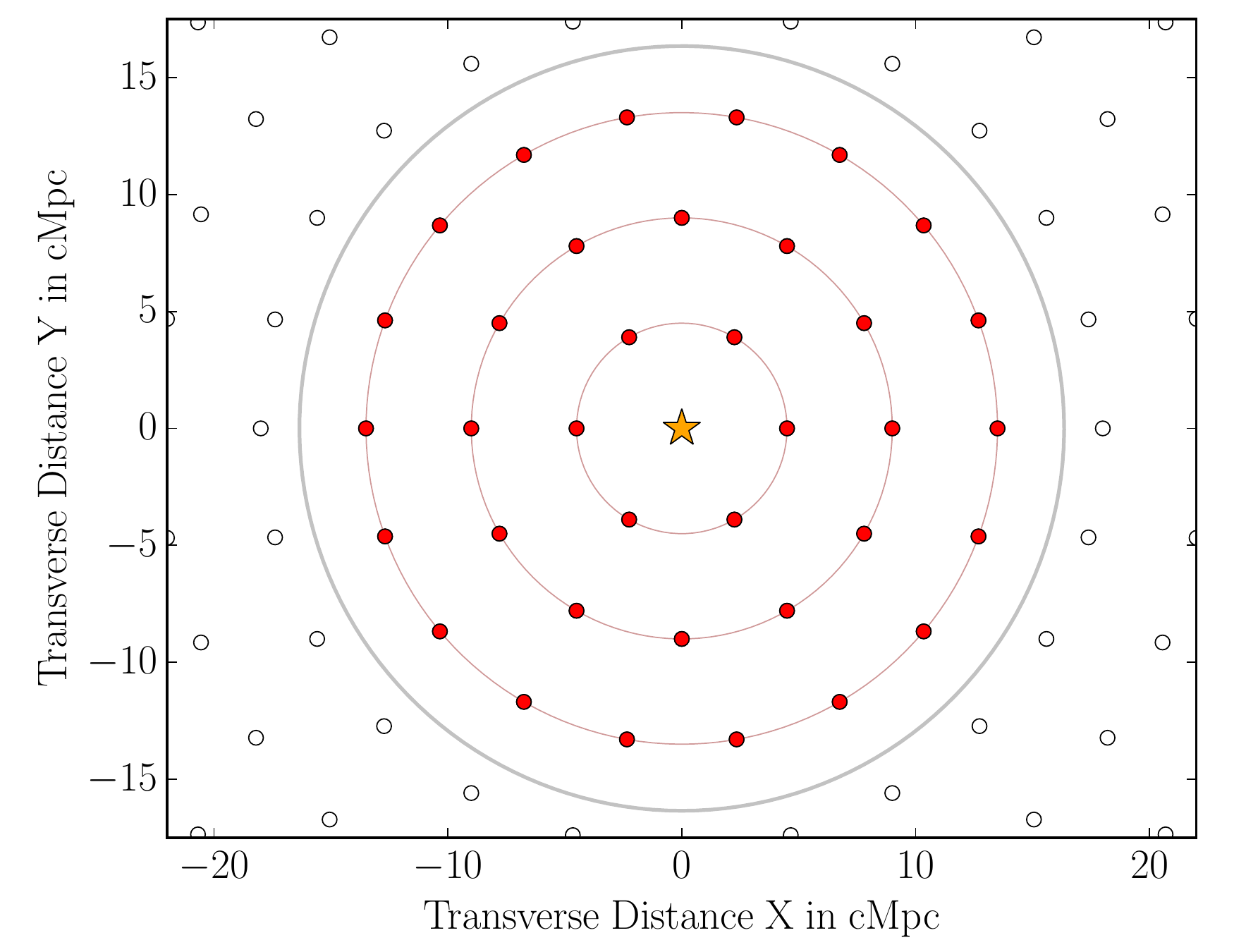}
 \caption{The sightline pattern adopted within this study, seen along the line-of-sight with the foreground quasar located in the center (orange star). The background sightlines (red points) are arranged on concentric circles with multiples of six sightlines per circle. An average  sightline separation of $D_\mathrm{SL}=4.5\cMpc$ is shown but the whole pattern might be rescaled to the desired sightline density. Sightlines outside the $16^\prime$ diameter FoV (gray circle) are discarded (open points).
 }
 \label{Fig:SightlinePattern}
\end{figure}

Given the selected halos, skewers are extracted along one of the grid axes using the sightline pattern illustrated in Figure~\ref{Fig:SightlinePattern}. The whole pattern is rescaled, i.e. stretched or compressed in radial direction, to match the desired average sightline density (see Figure~\ref{Fig:SightlineDensity}) and sightlines with a transverse separation larger than the adopted field-of-view (usually $16^\prime$ diameter) are discarded. Along the line-of-sight (i.e. velocity space), we center the skewers on the halo position in redshift space, taking the peculiar velocity of the halo into account. With the redshift of the foreground quasar as the origin, we assign individual redshifts to every pixel along the skewers. 

To better represent redshift evolution of the density field, we rescale the density of each pixel according to
\begin{equation}
\rho(z) = \rho_\mathrm{sim}  \times \left( \frac{z+1}{z_\mathrm{sim}+1} \right) ^3\:.
\end{equation}
We convert from simulated cosmic baryon density to hydrogen number density $n_\mathrm{H}$ using the primordial abundances of 76\% \citep{Coc2015}.  
The temperature and velocity field are taken directly from the simulation box without any change.

\subsection{Background Photoionization Rates}
\label{Sec:UV_Background}

Apart from the ionizing radiation of the foreground quasar, we adopt a spatially uniform UV background. \citet{Onorbe2017} presented an empirical relation for the cosmic mean transmitted \ion{H}{i} flux $\langle F_\mathrm{HI} \rangle$ fitted to existing measurements \citep{Fan2006, Becker2007, Kirkman2007, Faucher-Giguere2008, Becker2013a} of the form
\begin{equation}
 \tau_\mathrm{HI} = 0.00126 \times  e^{ 3.294 \, \times \sqrt{z} }
\end{equation}
where $\tau_\mathrm{HI} = \ln{ \langle F_\mathrm{HI} \rangle }$ denotes the effective optical depth and $z$ the redshift.
For simulation snapshot available at $z= 2.2, 2.5, 3.0, 3.5, 4.0, 4.5, 5.0$, 
we determine the mean transmission in a large set of random skewers and iteratively adjust the homogeneous \ion{H}{i}~UV background until the mean transmission matches the relation from \citet{Onorbe2017}. We interpolate these $\GuvbH$ values determined for the fixed redshifts using a cubic spline to obtain a smooth function $\GuvbH(z)$. This allows us to assign to each pixel the appropriate \ion{H}{i} UV background matched to its redshift.

\subsection{Foreground Quasar Photoionization Rates}
\label{Sec:Quasar_Ionization_Model}

Based on the the assumed $M_{1450}$ magnitude of the foreground quasars and assuming the \citet{Lusso2015} template for the spectral energy distribution of the quasars, we compute the quasar luminosity $L_\nu$, and from this the flux density $F_\nu$ at the background sightline according to
\begin{equation}
F_\nu = L_\nu \; \frac{1}{4\pi\:D_\mathrm{prop}^2}\;e^{-\frac{D_\mathrm{prop}}{\lambda_\mathrm{mfp}}}\;.
\label{Eq:Fnu_QSO}
\end{equation}
Here, $D_\mathrm{prop}$ denotes the proper 3-D distance from the foreground quasar to a specific position at the background sightline and $\lambda_\mathrm{mfp}$ is the mean free path to \ion{H}{i} ionizing photons.
We ignore IGM absorption by setting $\lambda_\mathrm{mfp}=\infty$, which is appropriate for the redshifts and scales we probe.\footnote{For $z_\mathrm{QSO} \gtrsim 5$, $\lambda_\mathrm{mfp}$ becomes comparable to the size of the proximity region \citep{Worseck2014} and IGM absorption can no longer be neglected.}

We assume the quasar spectra to be of power-law shape $f_\nu \propto \nu^\alpha$ with slope $\alpha = -1.7$ beyond $912\,\mathrm{\AA}$ \citep{Lusso2015} and  for simplicity also treat the spectral dependence of the hydrogen ionization cross-section as a power-law  of form $\sigma_{\nu} \propto (\nu/\nu_0)^{-3}$.
Using the cross-sections at the ionizing-edges $\sigma_0$ from \citet{Verner1996b} leads to the \ion{H}{i} photoionization rate
\begin{equation}
 \Gamma_\mathrm{QSO}^\mathrm{HI} = \int_{\nu_o^\mathrm{HI}}^\infty \frac{ F_\nu \; \sigma_{\nu}^\mathrm{HI} }{ \mathit{h}_\mathrm{P} \, \nu } \: d\nu\; \approx \frac{ F_{\nu_o^\mathrm{HI}} \; \sigma_0^\mathrm{\,HI} }{\mathit{h_\mathrm{P}} \; (3 - \alpha)}, \label{Eq:Q_QSO}
\end{equation}
where $\mathit{h}_\mathrm{P}$ denotes Planck's constant and $\nu_0^\mathrm{HI}$ is the frequency of the \ion{H}{i} ionization edge, i.e. $\mathrm{912\,\AA}$.

The photionization of \ion{He}{2} by the quasar might also have an effect on the thermal structure of the IGM \citep{Bolton2009, Bolton2010, Bolton2012}. However, proper treatment of this \textit{thermal proximity effect} would require radiative transfer calculations \citep{Meiksin2010, Khrykin2017} which is beyond the scope of this study.

\subsubsection{Finite Quasar Age}
\label{Sec:QuasarLifetime}

The above calculation implicitly assumes isotropic emission and a quasar luminosity $L_\nu$ that is constant for all times. In this section, we relax the latter assumption and compute which part of the background sightlines are illuminated by a foreground quasar of a given finite age.

A background sightline at transverse distance $R_\perp$ probes the foreground quasars emission at earlier times than the light we directly receive from the quasar \citep[see e.g.][]{Adelberger2004, Kirkman2008, Furlanetto2011, Schmidt2017, Schmidt2018}.
This arises from the fact that the geometric path length from the foreground quasar to a location along the background sightline, and from there to the observer (as probed in absorption by the background sightline) is longer compared to the direct path from the foreground quasar to Earth.
The total comoving path length is composed of the comoving distance from Earth to a point on the background sightline at redshift $z$ and from there to the foreground quasar. 
The sum of both can be converted to a redshift $z_\mathrm{initial}$ and corresponding lookback time $t_\mathrm{initial}$ at which the ionizing radiation from the foreground quasar had to be emitted.
\begin{figure}
 \centering
 \includegraphics[width=\linewidth]{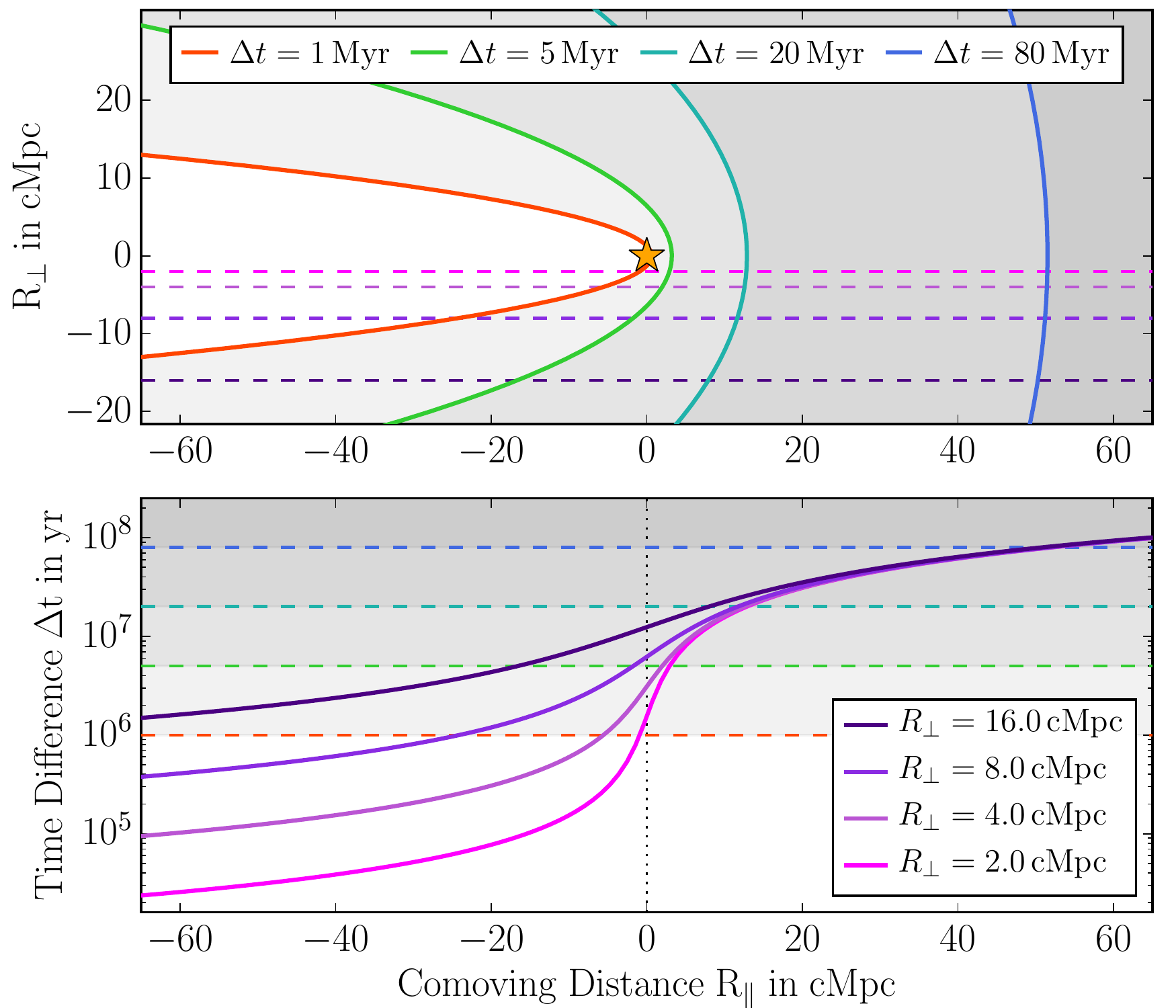}
 \caption{Visualization of the time difference $\Delta{}t$ in a slice around a $z_\mathrm{QSO}=3.2$ quasar (top panel).
 Curves of constant time difference appear as parabolas with the quasar located in the focal point. For $\tage= \Delta{}t$, only the region left of the corresponding $\Delta{}t$ curve appears to be illuminated.
 The bottom panel shows $\Delta{}t$ along four sightlines that pass by the quasar at different transverse separations $R_\perp$. 
 For $R_\parallel=0$, the time difference equals the transverse light crossing time. In front of the quasar ($R_\parallel<0$, $z<z_\mathrm{QSO}$), all sightlines probe smaller time differences, but the exact value has a strong dependence on the transverse separation. Behind the quasar ($R_\parallel>0$, $z>z_\mathrm{QSO}$), $\Delta{}t$ increases quickly with little dependence on $R_\perp$ and approaches $\Delta{}t = 2 \: R_\parallel \; \mathrm{c}^{-1}$.
 }
 \label{Fig:TimeRetardation}
\end{figure}
This lookback time at emission can be compared to the lookback time corresponding to the redshift of the foreground quasar $z_\mathrm{QSO}$. 
The difference is the additional time $\Delta{}t(z)$ it takes to first reach a certain point on the background sightline. 
If one neglects cosmic expansion (which we \emph{do not do} in
practice but do here for the sake of illustration), this simplifies to
\begin{equation}
\Delta{}t \approx \frac{a(z_\mathrm{QSO})}{c} \left( \sqrt{ R_\perp^2 + R_\parallel^2 } + R_\parallel \right),
\end{equation}
with $R_\perp$ and $R_\parallel$ denoting transverse and line-of-sight comoving separation from the quasar, $c$ the speed of light, and $a(z_\mathrm{QSO})$ the cosmic scale factor at the quasar redshift. Curves of constant time difference $\Delta{}t$ are thus parabolas.
In Figure~\ref{Fig:TimeRetardation} we give a detailed illustration of this behavior showing $\Delta{}t( \, R_\parallel \, )$ along four background sightlines that pass by a foreground quasar with transverse separations $R_\perp$ between $1\cMpc$ and $16\cMpc$.

\begin{figure}
 \centering
 \includegraphics[width=\linewidth]{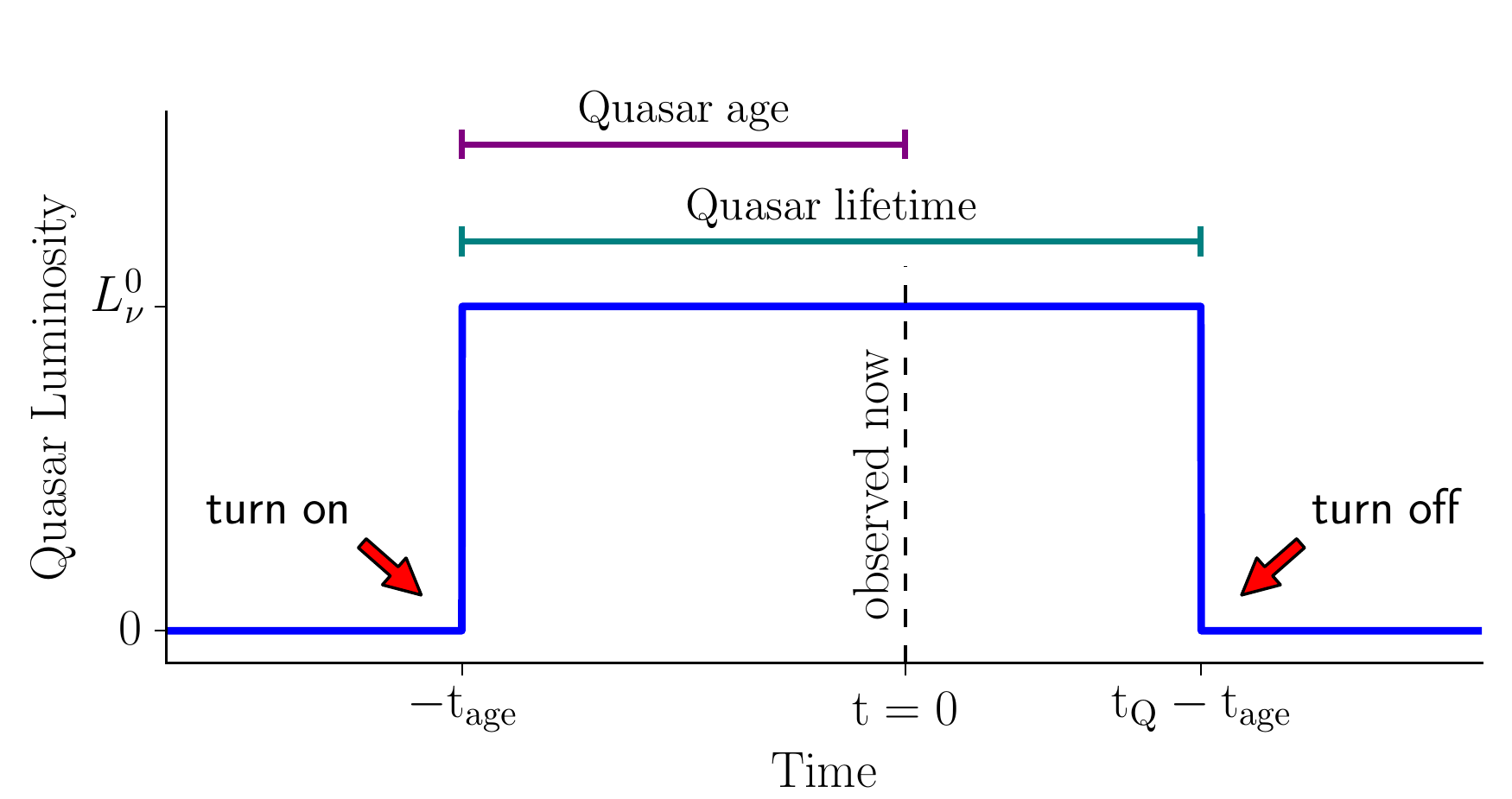}
 \caption{Visualization of the assumed quasar lightcurve.
 }
 \label{Fig:LightBulb}
\end{figure}

Since any point on a background sightline probes the quasar luminosity at a (different) earlier time than we observe the quasar today, we have to specify a quasar lightcurve $L_\nu(t)$. For this we assume a simple lightbulb model of the form
\begin{equation}
L_\nu(t) = L_\nu^0 \times  H( t + \tage ) \times H( \tQ - \tage - t), 
\end{equation}
in which $L_\nu^0 = L_\nu(\,t=0\,)$ is the currently observed quasar luminosity and the two $H(t)$ terms are Heaviside step functions to tun the quasar \textit{on} and \textit{off}. 
That is, we define the time at which the photons we observe today on Earth were emitted as $t=0$. In this model, the quasar \textit{turned on} at $t=-\tage$ and therefore we currently observe it at the age $\tage$. Its total lifetime is $\tQ$ and it will \textit{turn off} at time $t = \tQ - \tage$. Since we observe the quasar \textit{on} today, $\tQ > \tage$ and \textit{turn-off} will happen at some point $t>0$, i.e. \textit{in the future}. This diagram in Figure~\ref{Fig:LightBulb} illustrates these various times and aids visualization.

A certain point on a background sightline with time difference $\Delta{}t = \Delta{}t(\,R_\perp,R_\parallel\,)$ probes $L_\nu(-\Delta{}t)$. Therefore, points on a background sightline for which $\Delta{}t < \tTO$ appear for an observer on Earth to be illuminated by the foreground quasar. Since $\Delta{}t$ increases monotonically along the line of sight, i.e. with $R_\parallel$ or $z$ (see Figure~\ref{Fig:TimeRetardation}), all points at $R_\parallel$ higher than the dividing line where $\Delta{}t = \tTO$ appear not yet illuminated, simply because there was not enough time for the photons to reach these locations.

There is of course the possibility to assume different and probably more realistic quasar lightcurves. $\lya$ forest tomography should in principle be able to map the full emission history and able to constrain quasar variability over Myr timescales. For the moment however, we adopt the most basic lightbulb model and will explore the potential to constrain more complex quasar lightcurves in a future paper.

\subsubsection{Quasar Obscuration}
\label{Sec:QuasarObscuration}

Contrary to the established AGN unification paradigm \citep[e.g][]{Antonucci1993, Urri1995, Elvis2000, Netzer2015, Assef2013, Lusso2013, Padovani2017} that suggests that all quasars are obscured from some vantage points, for the purposes of this paper we assume that quasars emit isotropically.  The method we present here is capable of determining the quasar emission history as well as the quasar emission geometry. However, modeling the quasar radiation e.g. as a bi-conical emission pattern caused by an obscuring torus, adds a substantial amount of complexity to the analysis. Such a non-isotropic quasar emission model requires three more parameters, two angles $\theta$ and $\phi$ describing the orientation of the quasar emission cone or bi-cone, and an opening angle $\alpha$ setting the amount of obscuration or the opening angle of the emission cone(s).  We will demonstrate in a future paper how these parameters can also be inferred from tomographic $\lya$ observations, but for the sake of simplicity focus in this paper solely on quasar age and assume isotropic quasar emission.

\subsection{Ionization State of Hydrogen}

Based on the temperature $T$, velocity, and cosmic baryon density extracted from the Nyx simulation boxes, and adopting quasar and UV background photoionization rates as described above, we solve for the hydrogen ionization state.
At the redshifts we have in mind for this experiment ($z < 5$), hydrogen in the IGM is highly ionized by the metagalactic UV background \citep[e.g.][]{HaardtMadau2012, Planck2018VI}. We therefore assume ionization equilibrium, and that the IGM instantaneously adjusts to a change in the photoionization rate. This is well justified since the equilibration timescale for \ion{H}{i} at these redshifts is short, e.g. $\approx 10^{4}\yr$, compared to the timescales of interest. 

For calculating the \ion{H}{i} density $n_\mathrm{HI}$ we follow, like in \citet{Schmidt2018}, the approach described in \citet{Rahmati2013}.
We take the total ionization rate $\Gamma_\mathrm{tot}^\mathrm{HI}$ as the sum of photoionization $\Gamma_\mathrm{phot}^\mathrm{HI} = \GuvbH + \GqsoH$ and collisional ionization. For the photoionization we include the self-shielding prescription from \citet{Rahmati2013} in which the effective photoionization rate in high-density regions with $n_\mathrm{H} \gtrsim 5 \times 10^{-3}\,\mathrm{cm}^{-3}$ is substantially reduced.
For the collisional ionization we adopt the prescription by \citet{Theuns1998}.
We tie the fraction of helium in the \ion{He}{i} and \ion{He}{ii} states to the hydrogen ionization state by simply assuming $n_\mathrm{HeII}/n_\mathrm{He} = n_\mathrm{HII}/n_\mathrm{H}$.
Given the similar ionization energies, this is justified and a common assumption. 
We ignore the double ionization of helium for this study, i.e. assuming $n_\mathrm{HeIII} = 0$, since it adds unnecessary complications and has no substantial effect on the results.
We adopt the Case~A recombination coefficients from \citet{Storey1995}. This is appropriate since \ion{H}{i} is highly ionized and the IGM optically thin on the relevant scales.
With these inputs, the hydrogen ionized fraction can be easily computed.

\subsection{Computing Synthetic Spectra}

After determining $n_\mathrm{HI}$ along the skewers as stated above, the next step in our modeling procedure is to create synthetic spectra.
For each pixel along the skewers we compute an individual Voigt absorption line profile with appropriate strength, line width and velocity shift corresponding to the physical conditions in that pixel. Oscillator strengths are taken from \citet{Verner1996a}.
We benefit here from the high resolution of the \Nyx box ($36\ckpc$ or $\mathrm{2.8\,km\,s^{-1}}$) which is sufficient to resolve the $\lya$~forest ($\mathrm{\approx7.6\,km\,s^{-1}}$).
Redshift space distortions (peculiar velocities) are included by displacing the absorption profile with the line-of-sight velocity from the \Nyx simulation.
Thermal broadening is computed according to $\sigma_\mathrm{th} = \sqrt{ \frac{k_\mathrm{B} \, T}{m_\mathrm{H}} }$ for the Doppler broadening with $T$ denoting the gas temperature in a pixel and $m_\mathrm{H}$ the atomic mass of hydrogen. The Lorentzian scale parameter is based on the transition probability from \citet{Verner1996a}. 
The final transmission spectrum at a pixel in redshift space is the combination of all absorption profiles along a skewer.

\subsection{Error Forward Model}

Finally, we degrade these idealized spectra to account for observational effects by forward modeling finite spectral resolution, continuum errors, and photon-counting
and instrumental noise.

\subsubsection{Resolution}

We convolve the spectra with a Gaussian line spread function of appropriate width to simulate finite resolution of the 
spectrograph utilized, parametrized by its resolving power $R$. We also rebin the spectra in chunks of $1\cMpc$, which results in Nyquist sampling for resolving powers $R<2000$.

\subsubsection{Adding Noise to the Spectra}

We add random Gaussian noise to the rebinned spectra to mimick the desired $\rm{}S/N$ ratio. 
As already stated in \S\ref{Sec:SN_Requirement}, we specify the signal to noise per $R=1000$ or $300\,\mathrm{km\,s^{-1}}$ resolution element to keep the $\rm{}S/N$ measure independent of spectral resolution. The actual $\rm{}S/N$ per pixel will be lower by $\sqrt{N}$, with $N$ denoting the number of bins or pixels that sample a $300\,\mathrm{km\,s^{-1}}$ chunk of a spectrum.\footnote{For extremely low spectral resolution, $N$ might be $<1$ and the $\rm{}S/N$ per pixel in fact higher.}  
This procedure makes the exposure time required to reach a certain $\rm{}S/N_{1000}$ (approximately) independent of the adopted spectral resolution.

\subsubsection{Continuum Error}
\label{Sec:ContinuumError}

In addition to a random uncertainty per pixel due to photon-counting and instrumental
noise, there is a more systematic effect related to uncertainties in the continuum fitting. This effect could have a potentially severe impact on the measured $\lya$ forest transmission and therefore on the reconstruction of the quasar light echo. Thus, we forward model continuum fitting uncertainties following the description given in \citet{Krolewski2018}. Our final forward modeled $\lya$ transmission is therefore of the form
\begin{equation}
F_\mathrm{obs} = \frac{ F_\mathrm{true} + \delta_1 }{ 1 + \delta_2}
\label{Eq:Noise}
\end{equation}
where $F_\mathrm{true}$ is the simulated transmission while $\delta_1$ and $\delta_2$ are random Gaussian deviates with zero mean and standard deviations $\sigma_1$ and $\sigma_2$. 
The first term $\delta_1$ is related to the photon-counting and instrumental noise already mentioned above and drawn for each $1\cMpc$ bin individually. The standard deviation $\sigma_1$ is related to the desired $\rm{}S/N_{1000}$ and for $1\cMpc$ wide bins at $z_\mathrm{QSO}=3$ approximately  
\begin{equation}
\sigma_1 \approx \frac{1.98}{\rm{}S/N_\mathrm{1000}}.
\end{equation}

The second term $\delta_2$ corresponds to the continuum uncertainty and is identical for all bins along the same background sightline. Following \citet{Lee2012}, \citet{Krolewski2018} presented an empirical relation for the continuum uncertainty as function of the $\rm{}S/N$ of the data. Converting the given relation to the sampling used in this study we obtain 
\begin{equation}
\sigma_2 = \frac{0.409}{\rm{}S/N_{1000}} + 0.015 
\label{Eq:Noise_sigma2}
\end{equation}
which we will use throughout the paper to model continuum uncertainties.

\section{Simulated Observations of Quasar Light Echoes}
\label{Sec:SimulatedData}

\begin{figure*}[p]
 \centering
 \includegraphics[width=\linewidth]{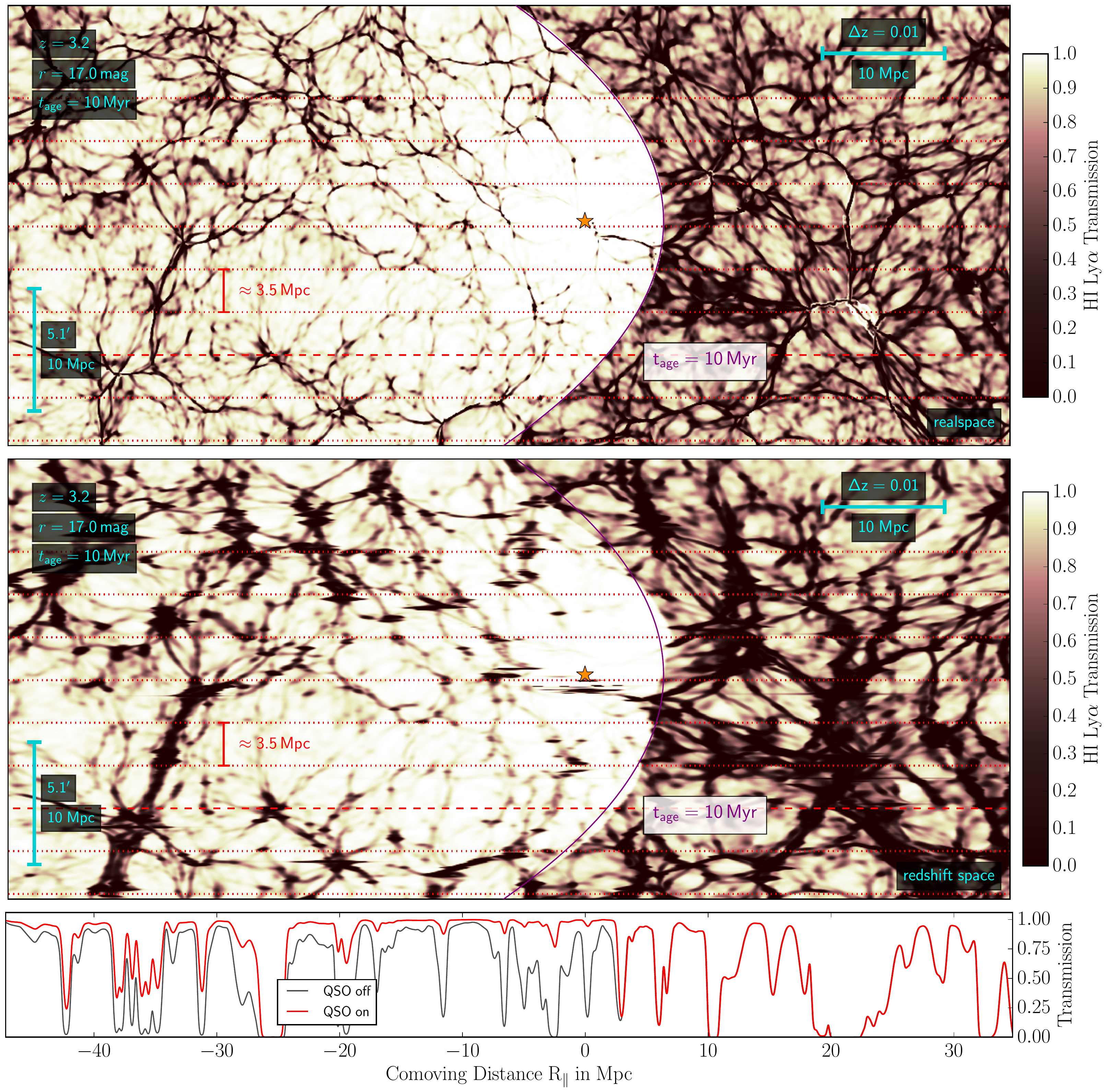}
 \caption{Visualization of the quasar proximity effect. The plot shows the \ion{H}{i} $\lya$ transmission in a slice through the simulation box. The snapshot of the hydrodynamical simulation was postprocessed with a photoionzation model of a $z_\mathrm{QSO}=3.2$ foreground quasar with an $r$-band magnitude of $r=17\Mag$, corresponding to $M_{1450}=-28.5\Mag$ that does shine for $10\Myr$.
 The top panel shows the situation in realspace, i.e. ignoring redshift space distortions and displaying directly the \ion{H}{i} transmission in each pixel without convolving with a line profile. Therefore, the region with enhanced transmission has exactly parabolic shape.
 The panel below displays the same but in redshift space. Here, peculiar velocities and the thermal broadening of the absorption lines was properly taken into account, which distorts the region of enhanced transmission and \textit{blurs} the overall picture.
 The bottom panel shows a computed mock spectrum along the red dashed sightline, once including the photoionization of the foreground quasar and once based solely on the photoionization of the metagalactic UV background. No binning was applied to the data and no noise or continuum errors added.
 A tomographic observation would probe the quasar proximity region with numerous background sightlines, on average spaced by e.g $\approx3.5\cMpc$ (red dotted lines).
 }
 \label{Fig:HI_TPE_Slice}
\end{figure*}

In what follows we briefly discuss the results of our simulations and try to build some intuition about the appearance of the \ion{H}{i} quasar proximity effect in 3D.
In Figure~\ref{Fig:HI_TPE_Slice} we show a two-dimensional slice through our simulation box. The quasar has a redshift $z_\mathrm{QSO}=3.2$ and an apparent magnitude of $r=17\Mag$, corresponding to $M_{1450}=-28.5\Mag$.  Without the foreground quasar, the IGM has a mean transmission of $\approx70\,\%$, however with substantial scatter ($\approx 15\,\%$ on $4\cMpc$ scales)
around this value due to cosmic density fluctuations. 
The ionizing radiation of the quasar increases the ionization state of hydrogen and pushes the \ion{H}{i} transmission to nearly $100\,\%$ in its immediate vicinity.  Even out to larger scales of several tens of megaparsec, a clear enhancement in the \ion{H}{i} $\lya$ forest transmission is visible.  The region of enhanced $\lya$ transmission has a clear boundary towards higher redshift (positive $R_\perp$, right in Figure~\ref{Fig:HI_TPE_Slice}), caused by the finite speed of light and the finite age of the quasar. The parabolic shaped surface corresponding to a path length
difference of $\Delta{}t = \tage = 10\Myr$ marks the boundary between the illuminated and non-illuminted region. 
For all points to the right (higher redshift, larger $R_\perp$), the quasar does not shine long enough to make these region appear illuminated for an observer on Earth.

However, while the sharp boundary of the quasar light echo exists in real space, this will not be the case in in redshift space.  A comparison of the top two panels in Figure~\ref{Fig:HI_TPE_Slice} illustrates the impact of redshift space distortions on the light echo structure. The large-scale velocity field displaces the apparent position of absorption features in a non-trivial way and causes the quasar light echo to be in some regions less and in some regions more extended in redshift space than in real space. This spatial distortion is related to the large-scale density field, since these density fluctuations source the bulk velocity flows. This can most easily be seen close to a large overdensity behind the quasar (at $R_\perp\approx12\cMpc$, upper half of the plot) that appears to drag the quasar light echo to the right towards the overdensity.

In principle, $\lya$ forest tomography delivers a map of the large-scale density structure, which might allow one to derive a model of the peculiar velocities and remove at least some part of these distortion from the tomographic map. However, due to the complex nature of the redshift space distortion, we do not undertake such reconstructions here. Peculiar velocities will therefore result in a form of correlated noise for the characterization of quasar light echoes. 

We quantify the amount of redshift space distortions in the map shown in Figure~\ref{Fig:HI_TPE_Slice} and find that the distribution of line-of-sight velocities is approximately Gaussian with a standard deviation of $125\,\mathrm{km\,s^{-1}}$. This corresponds to a FWHM of $295\,\mathrm{km\,s^{-1}}$ or $3.75\cMpc$ at $z_\mathrm{QSO}=3.2$. Although redshift space distortions are correlated and not directly comparable to the effects of finite spectral resolution, one can already see that the line-of-sight velocities correspond approximately to a resolving power $R=\frac{\lambda}{\Delta{}\lambda} \approx 1000$. Taking observations with substantially higher spectral resolution should therefore be of little benefit. However, we explore this dependence more thoroughly in \S\ref{Sec:ResResolvingPower}.

\begin{figure}
 \centering
 \includegraphics[width=\linewidth]{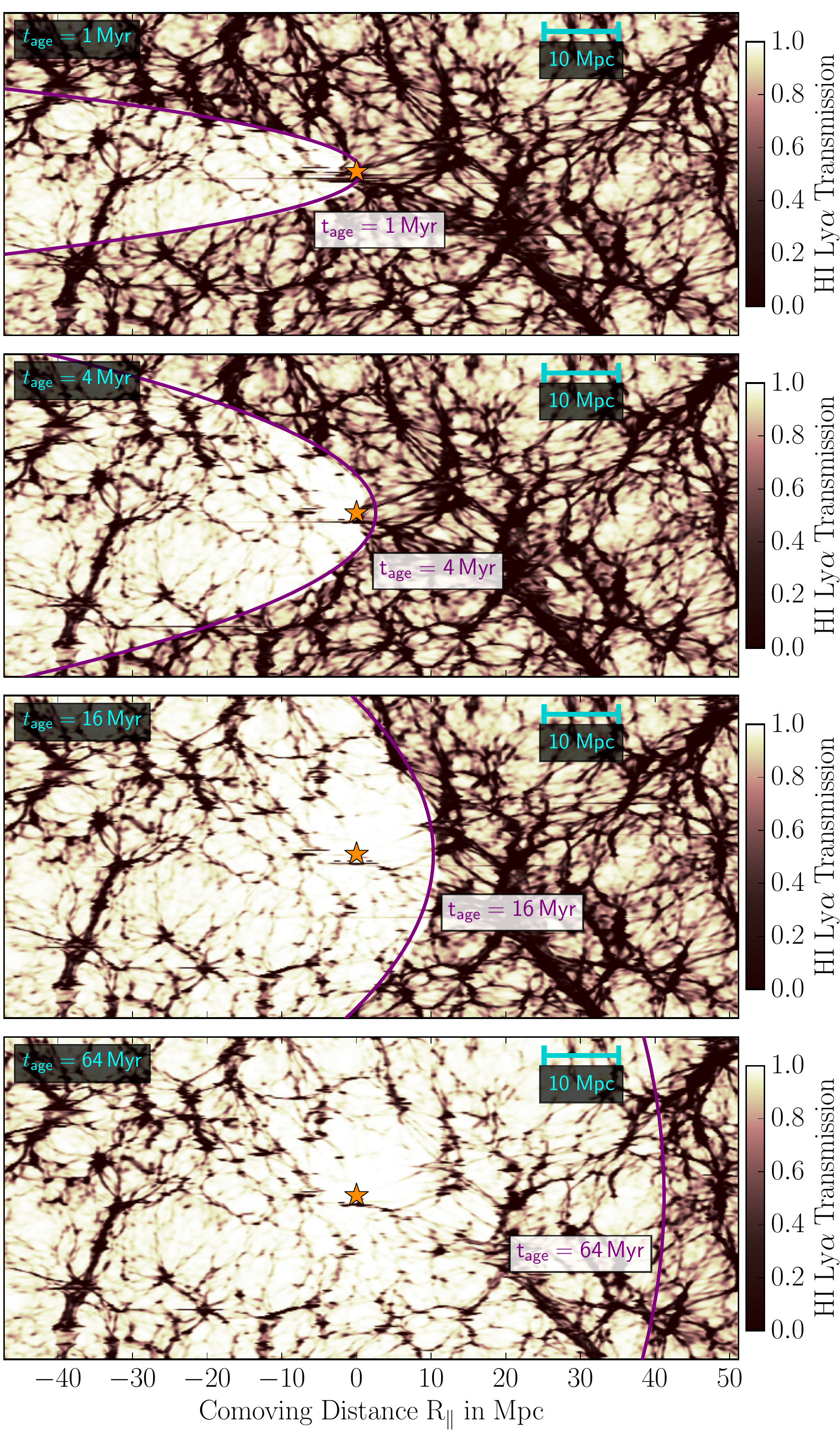}
 \caption{Appearence of the \ion{H}{i} proximity effet for four different quasar ages. The panels are similar to Figure~\ref{Fig:HI_TPE_Slice} and display redshiftspace. 
 The sequence of plots clearly visualizes how the illuminted region expands with increasing quasar age into the IGM. For $\tage=64\Myr$, the quasar light echo blends with the surrounding IGM, making it difficult to determine the extend of the proximity zone.
 }
 \label{Fig:HI_TPE_Sequence}
\end{figure}

In  Figure~\ref{Fig:HI_TPE_Sequence} we illustrate the time evolution of the proximity zone. Here, we again show the same slice through the simulation box as in Figure~\ref{Fig:HI_TPE_Slice} but compute the ionization state and Ly$\alpha$ forest transmission for different quasar ages between $\tage=1\Myr$ and $64\Myr$. This demonstrates how the quasar light echo expands into the IGM and how an increasingly large region around the quasar appears to be illuminated from Earth.

\subsection{IGM Transmission Statistics}
\label{Sec:TransmissionStatistic}

\begin{figure*}
 \centering
 \includegraphics[width=.49\linewidth]{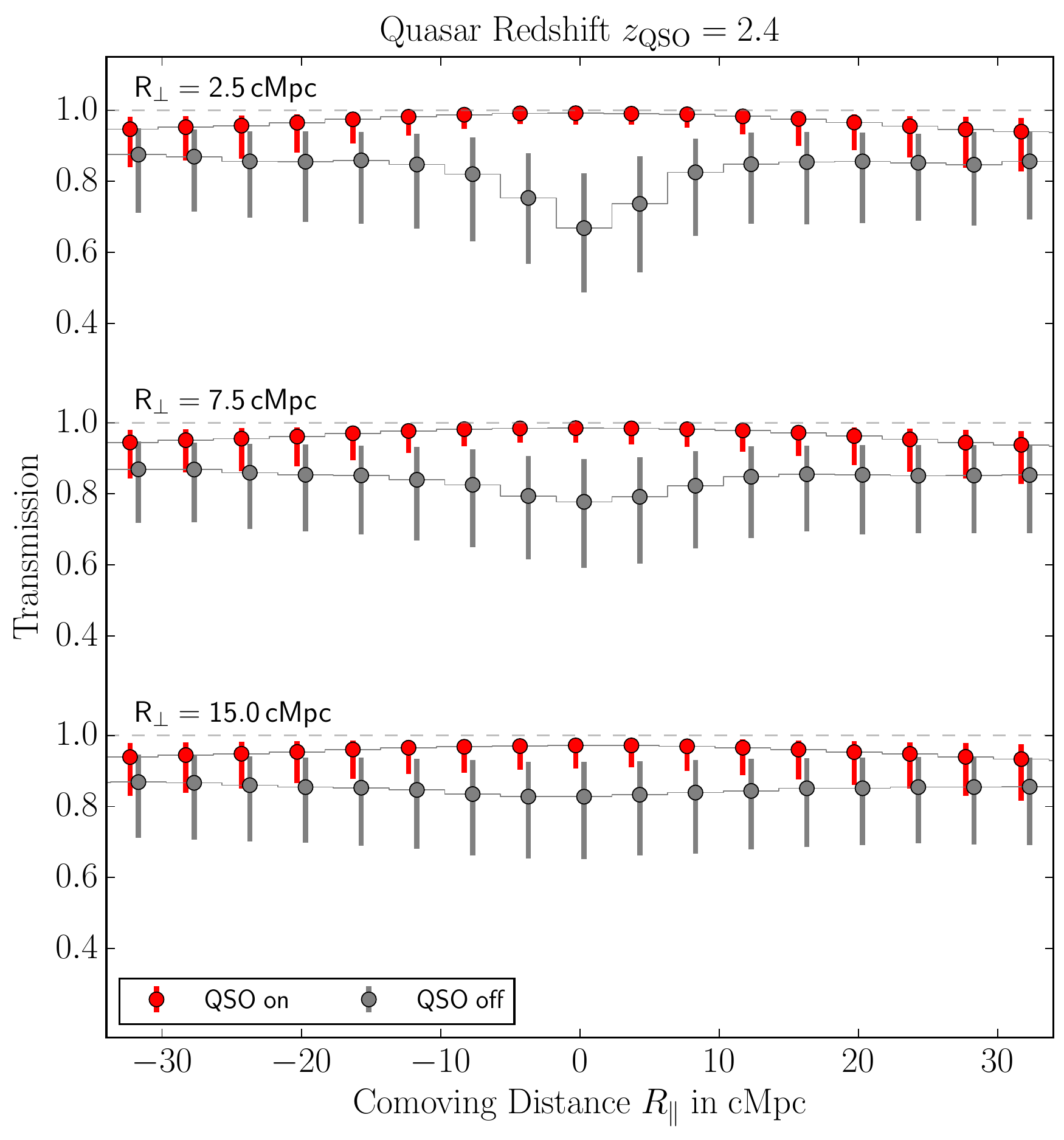}
 \hfill
 \includegraphics[width=.49\linewidth]{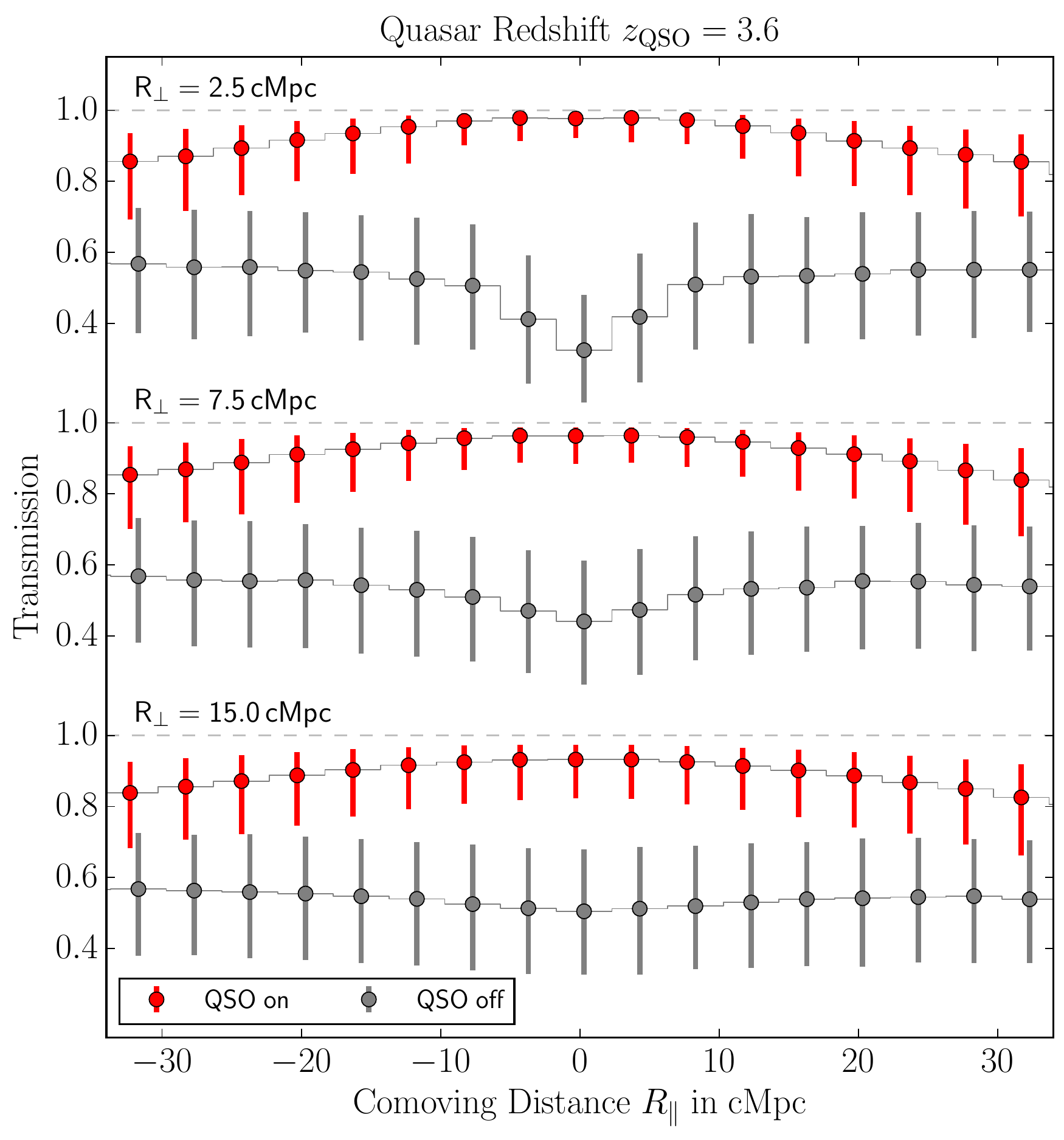}
 \caption{Average IGM transmission measured in $4\cMpc$ wide bins along background sightlines with transverse separations of $R_\perp = 2.5, 7.5$ and $15\cMpc$ from  an $M_{1450}=-28.7\Mag$ foreground quasar. The left panel shows the situation at $z_\mathrm{QSO}=2.4$, while the right displays $z_\mathrm{QSO}=3.6$. For each bin we show the case in which the sightlines are fully illuminated by the foreground quasar and the unilluminated case in which photoionization stems solely from the metagalactic UV background. 
 Close to the foreground quasar, excess absorption is visible due to the cosmic overdensity in which the quasar resides.
 Errorbars indicate the 16th -- 84th percentile interval of the expected IGM absorption in each bin, not including any observational noise.  
 Since the transmission is bounded at 100\,\%. the distributions in particular the for illuminated sightlines are highly skewed with the bulk of the distribution located at very high transmission values.
 }
 \label{Fig:HI_FluxStatistic}
\end{figure*}

In Figures~\ref{Fig:HI_FluxStatistic} we illustrate the flux probability distribution of the pixels within our tomographic map. For three background sightlines at transverse separations of $R_\perp = 2.5$, $7.5$ and $15\cMpc$ we show the expected median transmission in $4\cMpc$ wide bins as well as the 16th--84th percentile region.  At low redshift ($z_\mathrm{QSO}=2.4$; Figure~\ref{Fig:HI_FluxStatistic} left) it is difficult to determine if a specific part of a background sightline is illuminated by the quasar.  Although the ionizing radiation from the foreground quasar dominates over the UV background, the expected median transmission for the illuminated case overlaps with the 84th percentile of the distribution for the unilluminated case in nearly all $R_\parallel$ bins. A significant difference between illuminated and unilluminated case exists only very close to the quasar ($|R_\parallel|< 10 \cMpc$ and $R_\perp=2.5\cMpc$ or $7.5\cMpc$) where the cosmic overdensity of the host halo causes in absence of the quasars ionizing radiation excess absorption whereas the transmission is $\simeq 1$ if the quasar illuminates this region.
Therefore, it will be very challenging to detect and characterize quasar light echoes at such low redshifts.

The situation substantially improves at higher redshift. At $z_\mathrm{QSO}=3.6$ (Figure~\ref{Fig:HI_FluxStatistic} right), the mean IGM transmission drops below $60\,\%$ while the expected transmission in the illuminated case remains basically the same and still reaches $>85\,\%$ for $|R_\parallel|<30\cMpc$. Given that the IGM scatter only slightly increases, the 16th--84th percentile regions of the two distributions have basically no overlap and one can in principle for any individual bin along a background sightline infer with high confidence if that bin is illuminated by the foreground quasar. This higher contrast between illuminated and unilluminated parts of the IGM at high redshift  makes a detection of the proximity effect far easier then at low redshift, however at the expense of reduced background sightline density (see Figure~\ref{Fig:SightlineDensity}).

\section{Inferring Parameters from Quasar Light Echoes}
\label{Sec:LikelihoodComputation}

Inferring quasar properties from tomographic observations requires a statistical comparison of the observed data to a set of models. The analysis scheme we develop for this task has to be able to cope with several challenges. 

First, it has to combine and jointly fit the information from all transmission measurements along all background sightlines. This can, depending on the quasar redshift and limiting magnitude of the observations, result in up to twenty thousand individual measurements per tomographic map.

Second, the statistical analysis has to keep track of the correlations between individual measurements and cope with the intrinsically non-Gaussian transmission distributions in individual bins. This is of particular concern in the illuminated parts of the background sightlines which have transmissions close to 100\,\% (see Figure~\ref{Fig:HI_FluxStatistic}). Also, the measured IGM transmission has a very non-linear behavior with respect to the model parameters. Roughly speaking, it switches for a given spatial position between two binary states, depending on whether this part of the sightline is illuminated by the foreground quasar or not.

Third, we require the analysis to be fully Bayesian. This will allow us to deduce posterior probabilities for the inferred parameters and to determine meaningful confidence intervals. Additionally, it is desirable to have an analysis method that can handle degeneracies between model parameters. This is technically not necessary for the current paper since focus on inferring a single parameter, the quasar age, but our goal for the future is to generalize the modeling and inference to enable joint fits for quasar age, quasar orientation, and quasar obscuration. Particularly for short quasar age and high obscuration, one expects significant degeneracies between parameters.

Our approach to solve the issues outlined above using a set of dedicated models of the proximity effect and a Bayesian approach employing so called \textit{likelihood free inference}.

\subsection{Transmission Probability Distribution Functions}

For a given foreground quasar ($z_\mathrm{QSO}$, $M_{1450}$), proximity effect models are created as described in \S\ref{Sec:HydroSims}, based on outputs of a cosmological hydrodynamical simulation and postprocessed with a quasar photoionization model. We compute a model grid in $\tage$ that spans from $0.7\Myr$ to $128\Myr$ with 16 logarithmically spaced values of $\tage$.  For each of these models, we create 100 different model realizations which have the same quasar properties (i.e. $\tage$, $M_{1450}$, $z_\mathrm{QSO}$) and employ the same sightline pattern (illustrated in Figure~\ref{Fig:SightlinePattern}) but are centered on different host halos and therefore have different IGM density, velocity and temperature structure. These 100 independent model realizations are necessary to properly characterize the stochasticity of the IGM absorption. We forward model observational effects like finite spectral resolution, signal to noise ratio and continuum fitting errors to make the model outputs directly comparable to observed spectra. Each spectrum extends for $\pm65\cMpc$ around the foreground quasars position and we bin the spectra in chunks of $1\cMpc$ length.

For each chunk we obtain a kernel density estimate (KDE) of the probability distribution (PDF) of the transmission values in the chunk based on the 100 independent realizations of that model. This results in smooth functions, $p\,(\, F_{n,m} |\,\THETA \,)$, that describe the probability for measuring an IGM transmission $F_{n,m}$ in bin $n, m$, given the model parameter $\THETA$.  Here, the index $n$ identifies the different background sightlines within the sightline pattern while $m$ denotes a certain chunk along the sightline.  As long as we only focus on the quasar age, the parameter vector $\THETA$ has only one component $\THETA =\{ \tage \}$.  However, we still denote it as a vector since this approach allows for easy and straight-forward generalization of the inference to additional parameters like quasar obscuration or orientation. Following this, each model of the \ion{H}{i} proximity effect is fully described by a set of transmission probability functions $\mathcal{M} ^{\THETA} = \{ p\,(\,F_{n,m} |\,\THETA \,) \}$.

\subsection{Likelihood Computation}
\label{Sec:Likelihood}

Due to the high dimensionality of the observable, $\boldsymbol{F} = \{F_{n,m}\}$, i.e. several thousand individual transmission measurements, determining the likelihood $\mathcal{L} = p\,(\,\boldsymbol{F} \,|\, \THETA \,)$ poses a very challenging task.  The usual approach of approximating the likelihood as a multi-variate Gaussian is inadequate for our problem since the individual transmission PDFs are not well described by Gaussians.  In addition, determining the significant correlations between all the elements of $\{F_{n,m}\}$ would likely require an excessive number of model realizations. Therefore, we follow a likelihood free approach for which we never have to actually write down an analytic form for the likelihood function.

In \citet{Schmidt2018} we solved a similar problem. There however, the dimensionality of the problem was low (at most three transmission measurements and two model parameters) which made it feasible to simply map the full likelihood space by brute-force sampling. An approach like this would be completely impossible for our current case. A fully Bayesian treatment is only achievable if the dimensionality of the problem can be drastically reduced. We do this by first computing a so called \textit{pseudo-likelihood} which acts in many ways like a proper likelihood except that it ignores correlations between the $\{F_{n,m}\}$. In a second step, we then map this \textit{pseudo-likelihood} to determine a proper posterior probability distribution. Our approach is in many ways inspired by \citet{Alsing2018} and \cite{Davies2018c}, but is customized to the problem at hand and is in many ways different from either of these strategies.

For a given set of observed IGM transmissions $\boldsymbol{F}$, we define the \textit{pseudo-likelihood} $\mathcal{L}^\prime$  as
\begin{equation}
  \mathcal{L}^\prime \,(\, \boldsymbol{F} \,|\, \THETA \,) = \prod_{n,m}  \;
  \{ p\,(\,F_{n,m} |\,\THETA \,) \}.
  \label{Eqn:PseudoLikelihood}
\end{equation}
Therefore, we evaluate the transmission probability function of each chunk at the observed transmission level $F_{n,m}$, and compute the product of these probabilities. If the individual bins were uncorrelated this pseudo-likelihood would indeed represent the true likelihood function.  However, since this is in general not true, the pseudo-likelihood will not result in the correct parameter uncertainties and may also produce biased results.

We therefore only use it to find a parameter vector $\hat\THETA$ that maximizes this \textit{pseudo-likelihood} $\mathcal{L}^\prime$. This acts as a data compression and reduces the dimensionality of the data (up to several thousand) to the dimensionality of the model parameter space (a few or in our current case only one $t_{\rm age}$). During this process, some information might be lost, however, $\hat{\THETA}$ does retain the essence of information contained in the data and an approach like this has been proven to be a rather efficient and successful data compression algorithm \citep{Davies2018c}.

Since we only have models available for a set of discrete $\THETA$ values, the maximization process described above requires interpolation between models. We do this by interpolating the logartihm of the transmission probabilities, $\log \{ \, p\,(\,F_{n,m}|\,\THETA \,) \, \}$, evaluated at the specific observed value of $\,F_{n,m}^\mathrm{obs}$, using a simple quadratic interpolation scheme.

Compressing the dimensionality of the observable to the dimensionality of the model as described above now allows a fully Bayesian treatment of the problem.This requires that we determine the mapping between our summary statistic $\hat{\THETA}$ and the true parameter vector $\THETA$, which means we require the conditional probability distribution $p(\, \THETA \,|\, \hat{\THETA} \,)$. The technical feasibility of this approach was presented by \citet{Alsing2018}, however based on a different data compression scheme.
Using Bayes' theorem, the conditional probability distribution $p(\, \THETA \,|\, \hat{\THETA} \,)$ can be written as
\begin{equation}
p(\, \THETA \,|\, \hat{\THETA} \,) = \frac{ p(\, \hat{\THETA} \,|\, \THETA \,) \: p(\, \THETA \,) }{ p(\, \hat{\THETA} \,) }. 
\end{equation}
Here $p( \, \THETA \,)$ is our prior on the model parameters, which we here assume to be flat in $\log(\, \tage \,)$, and $p(\, \hat{\THETA} \,)$ is the evidence, which is basically just a normalization.  Since we have a generative model that can create mock data realizations for a given parameter set $\THETA$, we can relatively easily determine $p(\, \hat{\THETA} \,|\, \THETA \,)$. The low dimensionality of the problem, only $1 + 1$, makes it computationally feasible to approximate this distribution by simply computing samples.  In practice, we draw 1600 parameters values $\THETA$ from the prior $p( \, \THETA \,)$, compute model realizations for these values that yield the mock measurements $\{ F_{n,m} \}$, and straightforwardly determine $\hat{\THETA}$ for each realization. 
The latter is done by evaluating the transmission probabilities $\mathcal{M} ^{\THETA} = \{ p\,(\, F_{n,m} |\,\THETA \,) \}$ for each of the 1600 mock realization $\{ F_{n,m} \}$, computing the  \textit{pseudo-likelihood} in Equation~\ref{Eqn:PseudoLikelihood}, and finding the value $\hat{\THETA}$ that maximizes it.
\label{Sec:Prior}

\begin{figure}
 \centering
 \includegraphics[width=\linewidth]{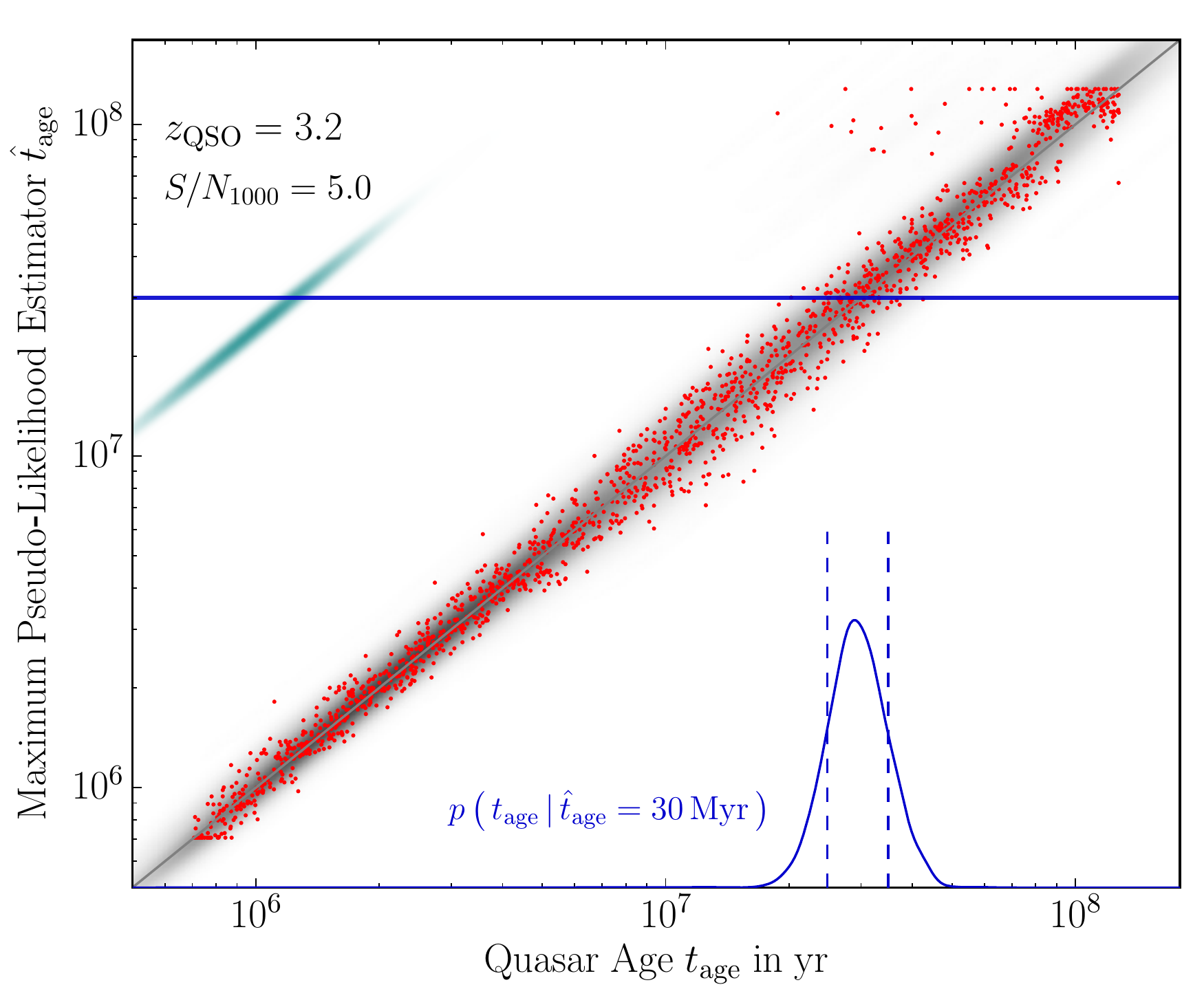}
 \caption{Mapping from maximum \textit{pseudo-likelihood} parameter $\hat{t}_\mathrm{age}$ to true posterior probabilities, based on 1600 model realizations (red points). These samples are convert to a smooth distribution (gray) by mean of KDE. The utilized kernel is shown in green in the top-left corner. 
 In blue, we illustrate the procedure to obtain posterior probabilities. 
 The $p(\, \hat{t}_\mathrm{age} \,|\, \tage \,) \: p(\, \tage \,)$ distribution is sliced at $\hat{t}_\mathrm{age} = \hat{t}_\mathrm{age}^\mathrm{obs}$ (here $30\Myr$) and re-normalized. The resulting posterior probability $p(\, \tage \, | \, \hat{t}_\mathrm{age}= \hat{t}_\mathrm{age}^\mathrm{obs} \,)$ is shown in the bottom right. Confidence intervals (16th and 84th percentile) are indicated with dashed lines. 
 }
 \label{Fig:Mapping}
\end{figure}

The resulting distribution of the 1600 samples in $(\, \hat{t}_\mathrm{age} \,|\, \tage \,)$ space is shown in Figure~\ref{Fig:Mapping}. As one can see, most samples are located relatively close to the 1:1 relation. Clearly, the $\{\hat{t}_\mathrm{age} \}= \hat{\THETA}$ that maximizes the \textit{pseudo-likelihood} $\mathcal{L}^\prime$ is a good proxy for the maximum of the true likelihood $\mathcal{L}$. The width of the distribution around the 1:1 relation is a measure for the width of the posterior and therefore the uncertainty in the parameter estimate. 

We note however, that in addition to this general scatter around the 1:1 relation, there are some outliers and artifacts in the distribution. For $z_\mathrm{QSO}<3$, where our method is less sensitive (see Figure~\ref{Fig:HI_FluxStatistic}), the \textit{pseudo-likelihood} maximizer has the tendency to run towards the upper boundary of the grid at $\hat{t}_\mathrm{tage}=128\Myr$. This effect happens mostly at low redshift (high mean IGM transmission) and long quasar ages. However, this issue quickly disappears for $z_\mathrm{QSO}\geq3$ and has in general very little impact on our analysis. 
Also, the $\mathcal{L}^\prime$ maximizer has a slight tendency to pick $\hat{t}_\mathrm{age}$ values that lie exactly on the model grid. This behavior is related to the relatively simple interpolation scheme we use for interpolation between the discrete models. Using a more sophisticated interpolator (e.g. Gaussian process interpolation, \citealt{Habib2007, Walther2018a, Walther2018b}.) would likely eliminate this issue. For now however, we monitor this behavior and ensure that, even if this artifact appears in our mapping procedure, it does not negatively impact our mapping. 

To derive proper posterior probabilities, we require a smooth and continuous version of the $p(\, \hat{t}_\mathrm{age} \,|\, \tage \,) \: p(\, \tage \,)$ distribution. We therefore apply a kernel density estimate to the 1600 computed samples which yields slightly better results than a Gaussian mixture model employed by \citet{Alsing2018}. We show the resulting interpolated distribution as gray shading in Figure~\ref{Fig:Mapping}. Note that the conditional probability $p(\, \hat{t}_\mathrm{age} \,|\, \tage \,)$ is nothing else than the joint probability $p(\, \hat{t}_\mathrm{age},\, \tage \,)$. Slicing this now smooth distribution at a specific value of $\hat{t}_\mathrm{age} = \hat{t}_\mathrm{age}^\mathrm{obs}$ and re-normalizing the result to unity, finally yields the proper posterior probability $p(\, \tage \, | \, \hat{t}_\mathrm{age}= \hat{t}_\mathrm{age}^\mathrm{obs} \,)$. 

Having set up our transmission probabilities  $\mathcal{M} ^{\THETA} = \{ p\,(\,F_{n,m} |\,\THETA \,) \}$. for a set of discrete values of $\THETA$ and the determining the poseterior $p(\, \THETA \,|\, \hat{\THETA} \,)$ via the procedure described above, we can for any simulated mock or observed IGM transmission $\boldsymbol{F} = \{F_{n,m}\}$ first determine the parameter vector $\hat{\THETA}$ that maximizes the \textit{pseudo-likelihood} $\mathcal{L}^\prime$ and then convert this into a proper posterior probability $p(\, \THETA \,|\, \boldsymbol F \,)$. 
In this way, we have a powerful and computationally feasible method to derive posterior probabilities in a fully Bayesian way that includes all effects related to correlations and non-Gaussianities without the requirement to write-down a likelihood function. It is simply based on the fact that we have a generative model capable of completely forward modeling
mock observations.

\section{Results}
\label{Sec:Results}

In the following, we will present which constrains $\lya$ forest tomography can impose on the quasar age.
To validate that our complex statistical analysis works as expected, we will create mock observations, analyze these with the fitting scheme described above and derive realistic posterior probabilities.
Furthermore, we will conduct an extensive parameter study to explore how the precision of the inferred $\tage$ depends on the properties of the foreground quasar, namely its redshift $z_\mathrm{QSO}$ and UV luminosity $M_{1450}$, as well as the observational setup. The latter is defined by the spectral resolution $R$, the covered field-of-view and the limiting magnitude $r_\mathrm{lim}$ for achieving the desired $\rm{}S/N_{1000}$ in a given exposure time. In consequence, these parameters also define the average sightline density $D_\mathrm{SL}$ and the number of sightlines $N_\mathrm{SL}$ that will probe the proximity region of the foreground quasar.
Determining the dependencies between these various parameters will be essential for choosing the optimal survey strategy for this project.

\subsection{Simulation Grid}

We create models of the 3D proximity effect for foreground quasar with redshifts ranging from $z_\mathrm{QSO}=2.4$ to $z_\mathrm{QSO}=5.0$ in steps of $\Delta{}z_\mathrm{QSO}=0.2$.
The fiducial setup assumes a limiting magnitude of $r_\mathrm{lim}=24.7\Mag$. 
For background sources of this brightness, existing multi-object spectrographs on 8--10\,m class telescopes (e.g. Keck/DEIMOS) should in good conditions achieve a $\rm{}S/N_{1000}=5$ in $10\,\mathrm{ks}$ (see Figure~\ref{Fig:ETC_Instruments} in \S\,\ref{Sec:SN_Requirement}). 
As shown in Figure~\ref{Fig:SightlineDensity}, the same limiting magnitude corresponds, depending on the foreground quasar redshifts, to different average sightline separations $D_\mathrm{SL}$. In Table~\ref{Tab:Sim_Grid_z}, we list the adopted values together with the number of background sightlines $N_\mathrm{SL}$ that fall into our $16'$ diameter FoV. We adopt the sightline pattern shown in Figure~\ref{Fig:SightlinePattern}. Table~\ref{Tab:Sim_Grid_z} also lists the adopted luminosity $M_{1450}$ of the brightest available quasars at these redshifts (see Figure~\ref{Fig:BrightestQuasars}).

For each of the models in Table~\ref{Tab:Sim_Grid_z}, we randomly select $N_\mathrm{Halo} = 100$ halos from the simulation box which define 100 independent sets of temperature, density and line-of-sight velocity along the background skewers.
Each sightline pattern created in the above way is then processed with our photoionization model as described in \S\ref{Sec:HydroSims}, assuming 16 different quasar ages $\tage=\{ 
0.7, \: 1.0, \: 1.4, \: \dots \: 128\}\Myr$, 
corresponding to $2^{\{-0.5, \: 0.0, \:  0.5, \: \dots \: 7.0 \}}\Myr$.
The background sightline spectra extend from $-65\cMpc < R_\parallel < 65\cMpc$ and are binned in chunks of $1\cMpc$ length, resulting in 131 independent transmission measurements per sightline. This defines for each model listed in Table~\ref{Tab:Sim_Grid_z} and each $\tage$ a set of transmission probabilities  $\mathcal{M} ^{\tage} = \{ p\,(\,F_{n,m} |\,\tage \,) \}$,

To realize the mapping from $\hat{t}_\mathrm{age}$ to $p(\, \tage \,|\, \hat{t}_\mathrm{age} \, )$ as shown in Figure~\ref{Fig:Mapping}, we compute a second set of models adopting the same parameters as listed in Table~\ref{Tab:Sim_Grid_z}. However, instead of simulating 16 discrete $\tage$ values for 100 IGM realizations, we compute 1600 realizations for which we draw $\tage$ randomly from our prior and pair it with a randomly selected halo above the minimum halo mass. As stated in \S\ref{Sec:Prior}, we adopt a prior that is flat in $\log( \tage )$.

For validation of our method, we create a set of models that act as mock observations.
For these we adopt quasar ages $\tage=\{1.0, \: 2.0, \: 4.5, \: 9.0, \: 20.0, \: 45.0, \: 80.0\} \Myr$ and choose random halos above the minimum halo mass. All other parameters are identical to the ones listed in Table~\ref{Tab:Sim_Grid_z}. For each quasar age, we compute 25 different IGM realizations. All models described above are post-processed to mimic a desired spectral resolution (usually $R=1000$), signal~to~noise ratio (e.g. $\rm{}S/N_{1000}=5$) and continuum uncertainties (see \S\ref{Sec:ContinuumError}).

After computing the various kinds of models, we have everything together for an end-to-end test of our method. 
For this, we apply the statistical analysis to the the mock observations and infer posterior probabilities for $\tage$, following the procedures described in \S\,\ref{Sec:LikelihoodComputation}. 

In this fitting process, we assume perfect knowledge of the foreground quasar redshift and do not include any uncertainties on this quantity. Given the overall expense of the tomographic observations and the extreme luminosity of the foreground quasar, it would in reality only add insignificant additional effort to obtain highly-precise redshift estimates e.g. from the [\ion{O}{iii}] line ($\Delta{}z\leq100\,\mathrm{km\,s^{-1}}$) using infrared spectroscopy or alternatively the CO or [\ion{C}{ii}] $158\,\um$ fine-structure lines ($\Delta{}z\leq50\,\mathrm{km\,s^{-1}}$) in the sub-mm regime.

We also assume perfect knowledge about the UV background and the quasars ionizing emissivity. This means in practice that the models we use for fitting have exactly the same mean transmission and quasar ionizing emissivity as the mock observations. The mean IGM transmission is relatively well known (to better than 2\%, \citealt{Becker2013a})
and the uncertainties probably dominated by the statistical fluctuations within the map. 
The quasar luminosity in the non-ionizing UV ($m_{1450}$) is directly observable and should be known to very high precision. The extrapolation from there to the ionizing regime adds some uncertainty due to the a~prior unknown quasar SED. However, variations in the quasar spectral slope relate to only moderate uncertainties in the ionizing flux, e.g. about $13\%$ based on \citet{Lusso2015}. In addition, we would preferentially select target quasars that have confirmed flux beyond the Lyman limit ($\lambda_\mathrm{rest}=912\,\mathrm{\AA}$), which could give additional constraints on the quasar SED. In the future, we anyway intend to include the (possibly time dependent) quasar luminosity in the analysis procedure and fit for the quasar ionizing flux.

\begin{deluxetable}{ccccccccc}
\centering
\tablecolumns{8}
\tablewidth{0pc}
\tablecaption{Parameter of the Main Simulation Grid.}
\tablehead{
\colhead{$z_\mathrm{QSO}$}	& \colhead{$M_{1450}$}	& \colhead{$r_\mathrm{lim}$}	& \colhead{$D_\mathrm{SL}$}	& \colhead{$\mathrm{FoV}$}	& \colhead{$N_\mathrm{SL}$}	&	\colhead{$R$}	&	\colhead{$\rm{}S/N_{1000}$}	\\
\colhead{}			& \colhead{$\Mag$}	& \colhead{$\Mag$}		& \colhead{$\cMpc$}		& \colhead{$'$}		& \colhead{}			&	\colhead{}	&	\colhead{}	
}
\startdata
$2.4$   & $-28.7$       & $24.7$        & $1.62$        & $16$  & $216$         & $1000$         & $5.0$         \\ 
$2.6$   & $-28.9$       & $24.7$        & $1.87$        & $16$  & $168$         & $1000$         & $5.0$         \\ 
$2.8$   & $-29.0$       & $24.7$        & $2.22$        & $16$  & $126$         & $1000$         & $5.0$         \\ 
$3.0$   & $-29.0$       & $24.7$        & $2.54$        & $16$  & $ 90$         & $1000$         & $5.0$         \\ 
$3.2$   & $-29.0$       & $24.7$        & $2.91$        & $16$  & $ 90$         & $1000$         & $5.0$         \\ 
$3.4$   & $-28.8$       & $24.7$        & $3.47$        & $16$  & $ 60$         & $1000$         & $5.0$         \\ 
$3.6$   & $-28.7$       & $24.7$        & $4.24$        & $16$  & $ 36$         & $1000$         & $5.0$         \\ 
$3.8$   & $-28.6$       & $24.7$        & $5.02$        & $16$  & $ 36$         & $1000$         & $5.0$         \\ 
$4.0$   & $-28.5$       & $24.7$        & $5.82$        & $16$  & $ 18$         & $1000$         & $5.0$         \\ 
$4.2$   & $-28.3$       & $24.7$        & $6.79$        & $16$  & $ 18$         & $1000$         & $5.0$         \\ 
$4.4$   & $-28.2$       & $24.7$        & $8.76$        & $16$  & $ 18$         & $1000$         & $5.0$         \\ 
$4.6$   & $-28.1$       & $24.7$        & $12.03$       & $16$  & $  6$         & $1000$         & $5.0$         \\ 
$4.8$   & $-27.9$       & $24.7$        & $14.31$       & $16$  & $  6$         & $1000$         & $5.0$         \\ 
$5.0$   & $-27.8$       & $24.7$        & $16.81$       & $16$  & $  6$         & $1000$         & $5.0$         
\enddata
\tablecomments{Parameters are chosen to keep the total observing time approximately constant, e.g. 3$\times$ $10000\,\mathrm{ks}$ with Keck/DEIMOS.
}  
\label{Tab:Sim_Grid_z}
\end{deluxetable}

\subsection{Example Posteriors}

\begin{figure*}
 \centering
 \includegraphics[width=\linewidth]{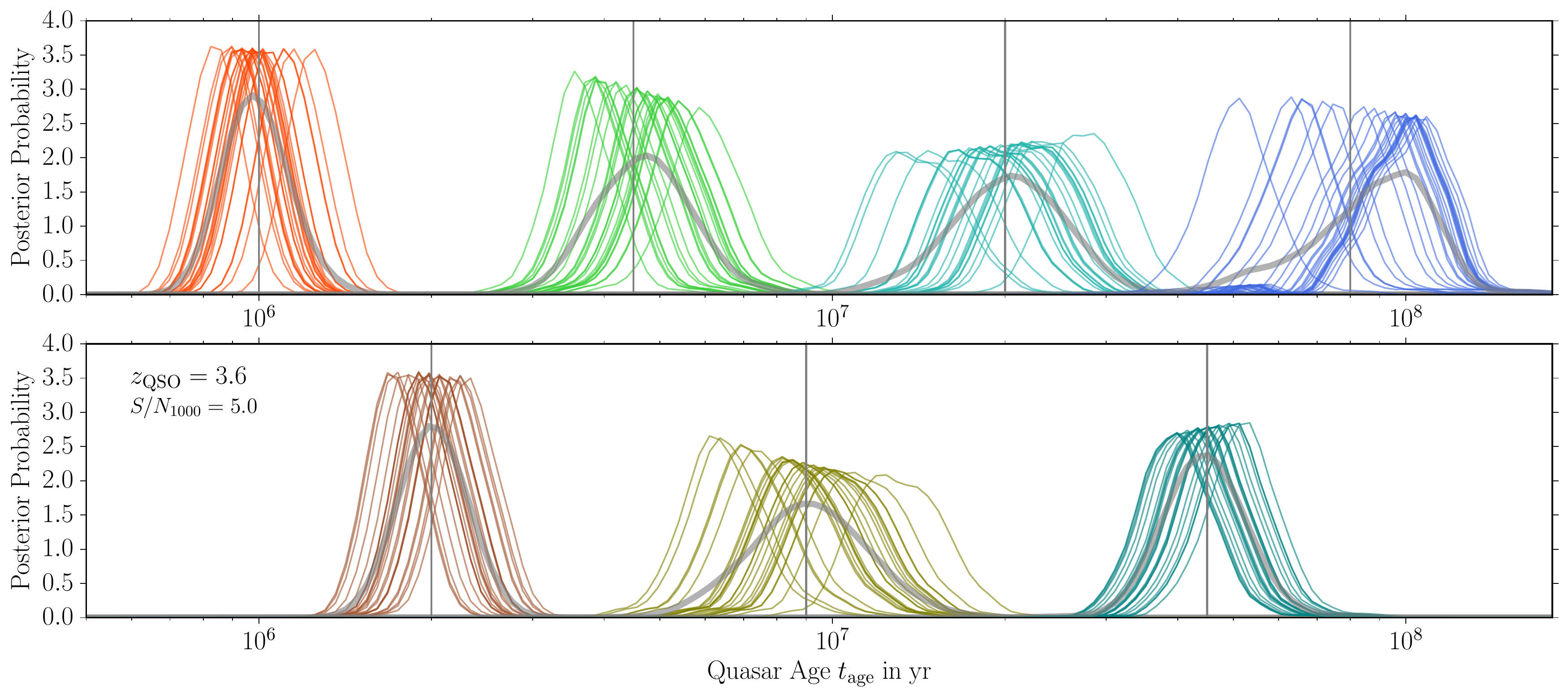}
 \caption{Posterior probability distributions for seven sets of mock observations with quasar ages (vertical lines) between $1.0$ and $80\Myr$. For each of the seven $\tage$, 25 independent realizations are shown (colored curves) and as a thick gray curve the average of the individual posterior probabilities. The adopted foreground quasar redshift is $z_\mathrm{QSO}=3.6$, spectral resolution is $R=1000$ and $\rm{}S/N_{1000}=5.0$. Further properties of the models are listed in Table~\ref{Tab:Sim_Grid_z}.\\
 }
 \label{Fig:PosteriorGallery}
\end{figure*}

A set of posterior probabilities derived in the described way for a model with
$z_\mathrm{QSO} = 3.6$ are shown in Figure~\ref{Fig:PosteriorGallery}. The figure clearly shows that our method works well and yields satisfying estimates for the quasar age.
The posterior probabilities are localized and in the right place. In most of the cases, the true $\tage$ value is well within the extent of the confidence interval (16th -- 84th percentile) and there are very few cases in which the derived $\tage$ estimate deviates substantially from the true value. Averaging the 25 individual posterior probabilities of each model gives an estimate of the achieved accuracy. This also shows that our method yields unbiased results. A slight exception from this might be the $80\Myr$ case, which approaches the highest quasar ages that can be constrained with $\lya$ forest tomography. At such long $\tage$, as can be seen in Figure~\ref{Fig:HI_TPE_Sequence}, the edge of the proximity region starts to blend smoothly with the IGM and the proximity region extends far beyond the adopted $16'$ FoV. The exact behavior depends however on the luminosity of the quasar and the mean transmission of the IGM and therefore redshift. Also note that the cut-off of the posterior probabilities towards high $\tage$ might to some degree be artificial since our model grid only extends up to $128\Myr$. At some point the results should be treated as lower limits.
In general, this test proves that $\lya$ forest tomography is indeed able to constrain quasar ages in a precise and reliable fashion.

Figure~\ref{Fig:PosteriorGallery} also reveals that, at least to first order, all seven models  show a similar width of the posterior probabilities, more or less independent of the true age of the quasar. Since the axis in Figure~\ref{Fig:PosteriorGallery} is scaled logarithmically, this translates to an approximately constant relative uncertainty $\frac{\Delta{}\tage}{\tage} \approx 20\,\%$. This general behavior can also be seen in the mapping from $\hat{t}_\mathrm{age}$ to $p(\, \tage \,|\, \hat{t}_\mathrm{age} \,)$ which defines the width of the derived posterior probabilities. The distribution shown in Figure~\ref{Fig:Mapping} has approximately constant width around the 1:1 relation.  However, there is some dependence on the quasar age which we discuss in more detail below.

\subsection{Dependence on $\tage$}

To quantify the precision of the derived $\tage$ estimate, we define the relative uncertainty as the 16th--84th percentile interval of the posterior probability parametrized as function of $\log( \tage )$ which yields  $\frac{\Delta{}\tage}{\tage}$.
\begin{figure}[bt]
 \centering
 \includegraphics[width=\linewidth]{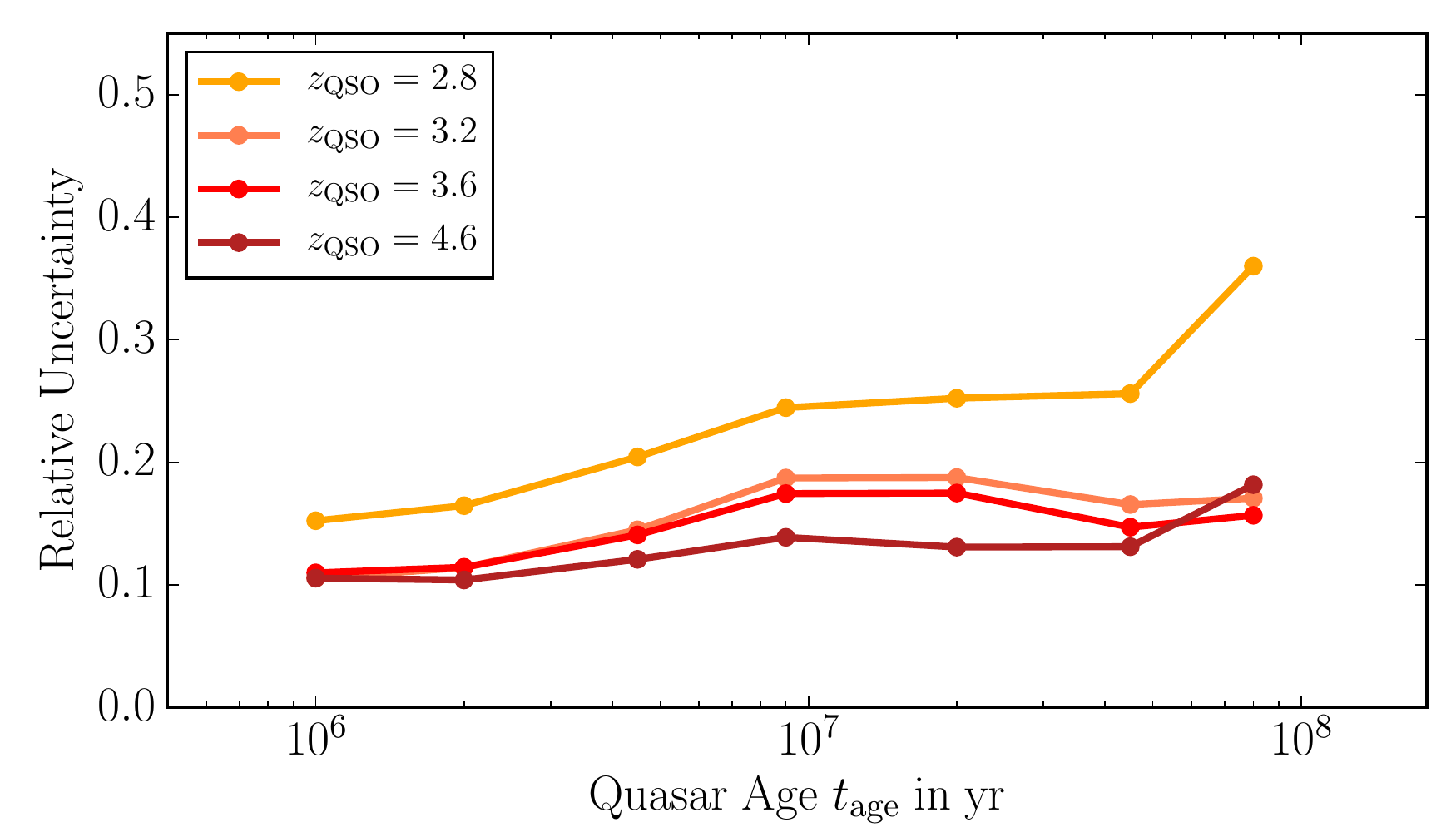}
 \caption{Dependence of the achieved precision on the adopted quasar age $\tage$. For each quasar age and three different quasar redshifts, the average precision of the 25 mock datasets is shown. Further model parameters are listed in Table~\ref{Tab:Sim_Grid_z}.
 }
 \label{Fig:Sequence_tage}
\end{figure}
Figure~\ref{Fig:Sequence_tage} shows this relative precision derived from individual posteriors averaged over the 25 realizations per model as function of quasar age.
As one can see, the highest relative precision around 10\% is achieved for young quasars ($\tage\approx1\Myr$). For very long quasar ages, similarly small uncertainties around 15\% can be reached. At intermediate ages around $\tage\approx10\Myr$, the precision is only around 20\%. However, this depends on the quasar redshift.
Quantitatively understanding the origin of the dependence of the precision on quasar age is not a trivial task, but we believe it is related to a combination of the smearing effect of redshift space
distortions and the geometry of the region illuminated by the quasar.

As shown, the quality of the derived constrains depends slightly on the adopted foreground quasar redshift. For low quasar redshifts, e.g the $z_\mathrm{QSO}=2.8$ case, the achieved precision is in general not as good as for $z_\mathrm{QSO}>3.0$ and deteriorates in particular for long quasar ages.

\subsection{Dependence on $z_\mathrm{QSO}$}

\begin{figure}
 \centering
 \includegraphics[width=\linewidth]{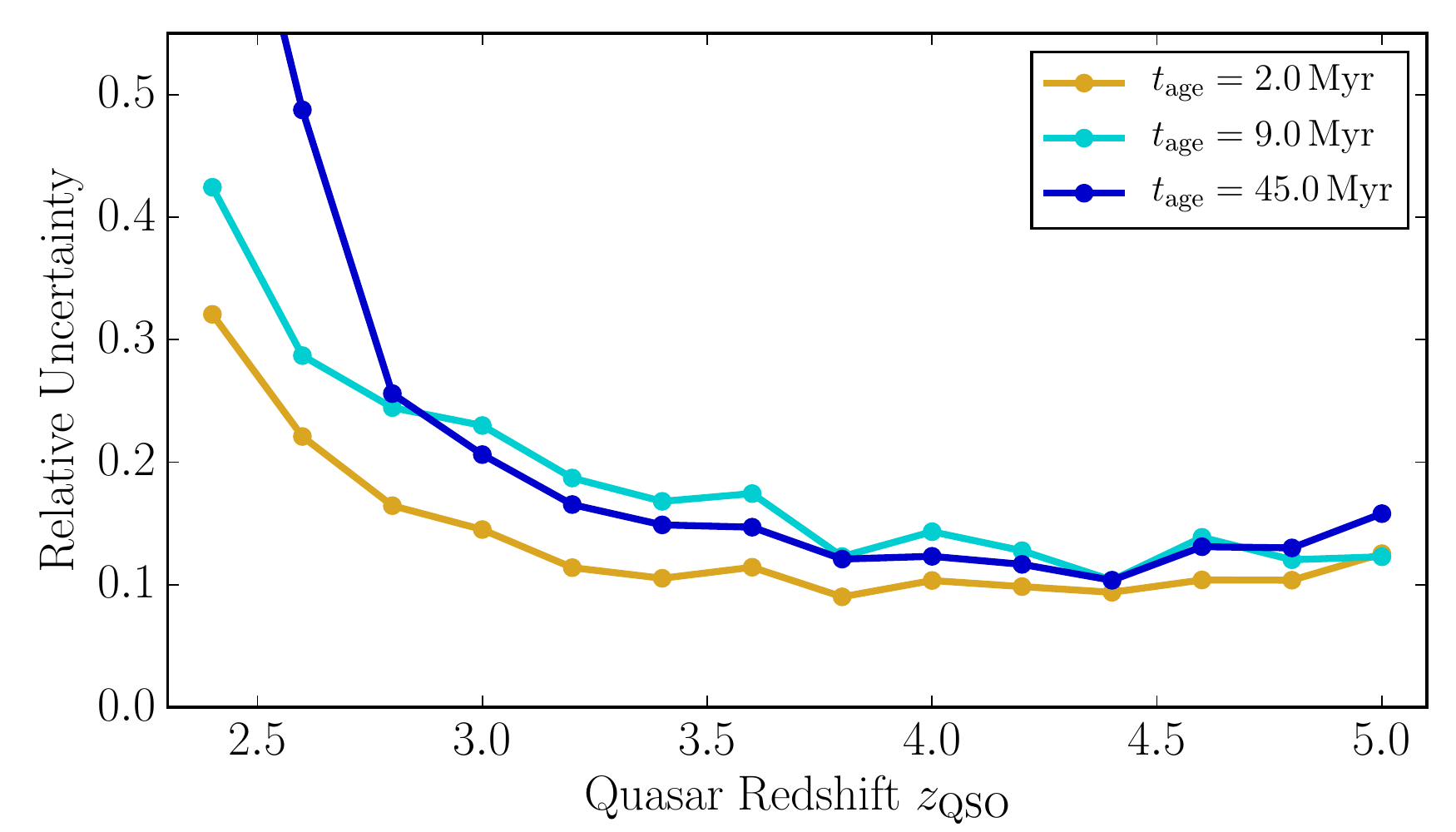}
 \caption{Dependence of the achieved precision on the adopted quasar redshift $z_\mathrm{QSO}$. For each quasar redshift and three different quasar ages, the average precision of the 25 mock datasets is shown. The chosen sightline separation and number of background sightlines within the $16'$ FoV corresponds to a constant limiting magnitude of $r_\mathrm{lim}=24.7\Mag$ at all redshifts. For $\rm{}S/N_{1000}=5$, this limiting magnitude should be achievable in $\approx 10\,\mathrm{ks}$ exposures. Further model parameters are listed in Table~\ref{Tab:Sim_Grid_z}.
 }
 \label{Fig:Sequence_zQSO}
\end{figure}

In Figure~\ref{Fig:Sequence_zQSO}, we explore the redshift dependence of our method in more detail. In what follows  we keep the limiting magnitude for achieving $\rm{}S/N_{1000}=5$ fixed at $r_\mathrm{lim}=24.7$, which results in an approximately constant exposure time around 10 ks (see Figures~\ref{Fig:ETC_Instruments} and \ref{Fig:SightlineDensity}). We also assume the same fixed FoV of $16'$.
As already outlined in \S\ref{Sec:OptimalQuasarRedshift}, one has to find a compromise between sampling the quasar light echo by many background sightlines at low $z_\mathrm{QSO}$ and the overall stronger proximity at higher redshift, owing to the lower average IGM transmission and thus increased contrast in the proximity zone (see Figure~\ref{Fig:HI_FluxStatistic}).

Figure~\ref{Fig:Sequence_zQSO} indicates that the latter effect clearly dominates. For $z_\mathrm{QSO}>3.3$, we achieve an age precision better than 20\,\%, nearly independent of redshift. Below this however, the $\tage$ precision degrades substantially, despite sampling the quasar proximity zone with up to 216 background sightlines at $z_\mathrm{QSO}=2.4$.
The deterioration is particularly dramatic for long quasar ages, where the transmission enhancement caused by the quasars ionizing radiation would have to be detected at large distances from the quasar, at which point the corresponding small transmission enhancement becomes indistinguishable from the average (high) IGM transmission given the relatively large stochastic fluctuations. 
As already discussed in \S\ref{Sec:TransmissionStatistic} and shown clearly in Figure~\ref{Fig:HI_FluxStatistic}, redshifts $z_\mathrm{QSO}\leq2.8$ are not well suited to map quasar light echoes and one should in general aim for higher redshift where the mean IGM transmission is lower.

At very high quasar redshift, ($z_\mathrm{QSO}\gtrsim4.5$) the average separation between sightlines becomes comparable to the size of our adopted field-of-view and the number of contributing background sightlines is rather low (see Table~\ref{Tab:Sim_Grid_z}).The rigid sightline pattern we adopt in our analysis (see Figure~\ref{Fig:SightlinePattern}) is not optimized for this regime.
This discretization in background sightline density causes undesired jumps and wiggles in the curves shown in Figure~\ref{Fig:Sequence_zQSO}, e.g at $z_\mathrm{QSO}=4.4$.%
\footnote{At $z_\mathrm{QSO}=4.4$, $D_\mathrm{SL}=8.8\cMpc$ and two rings of our pattern fit into the FoV, resulting in $N_\mathrm{SL}=18$.  At $z_\mathrm{QSO}=4.6$, the average sightline
separation is $D_\mathrm{SL}=12\cMpc$ and the quasar proximity region is only sampled by 6 background sightlines.}
For the low sightline density regime, a random placement of background sightlines in the FoV would clearly be more appropriate. 

A better assessment of the performance of our method at $z_\mathrm{QSO} \approx 5$ and beyond might therefore require a slightly different approach and a dedicated study.  This might reveal that the method can be pushed to even higher redshifts. However, one has to note that in this study we model the proximity effect with a rather simple model that has only one free parameter. Therefore, a single background sightline theoretically delivers sufficient information to fully constrain the model. However, our previous studies of the \ion{He}{ii} transverse proximity effect \citep{Schmidt2017, Schmidt2018} showed that measurements along single background sightlines, even in the case of low mean IGM transmission and high contrast, are often not sufficient to deliver strong and unique constraints on quasar properties when quasar age, obscuration and orientation effects are taken into account. We expect that in general a higher number of background sightlines is necessary to constrain models more complex than we consider here. We will address these questions in more detail in a future paper. 

\subsection{Dependence on S/N}

\begin{figure}
 \centering
 \includegraphics[width=\linewidth]{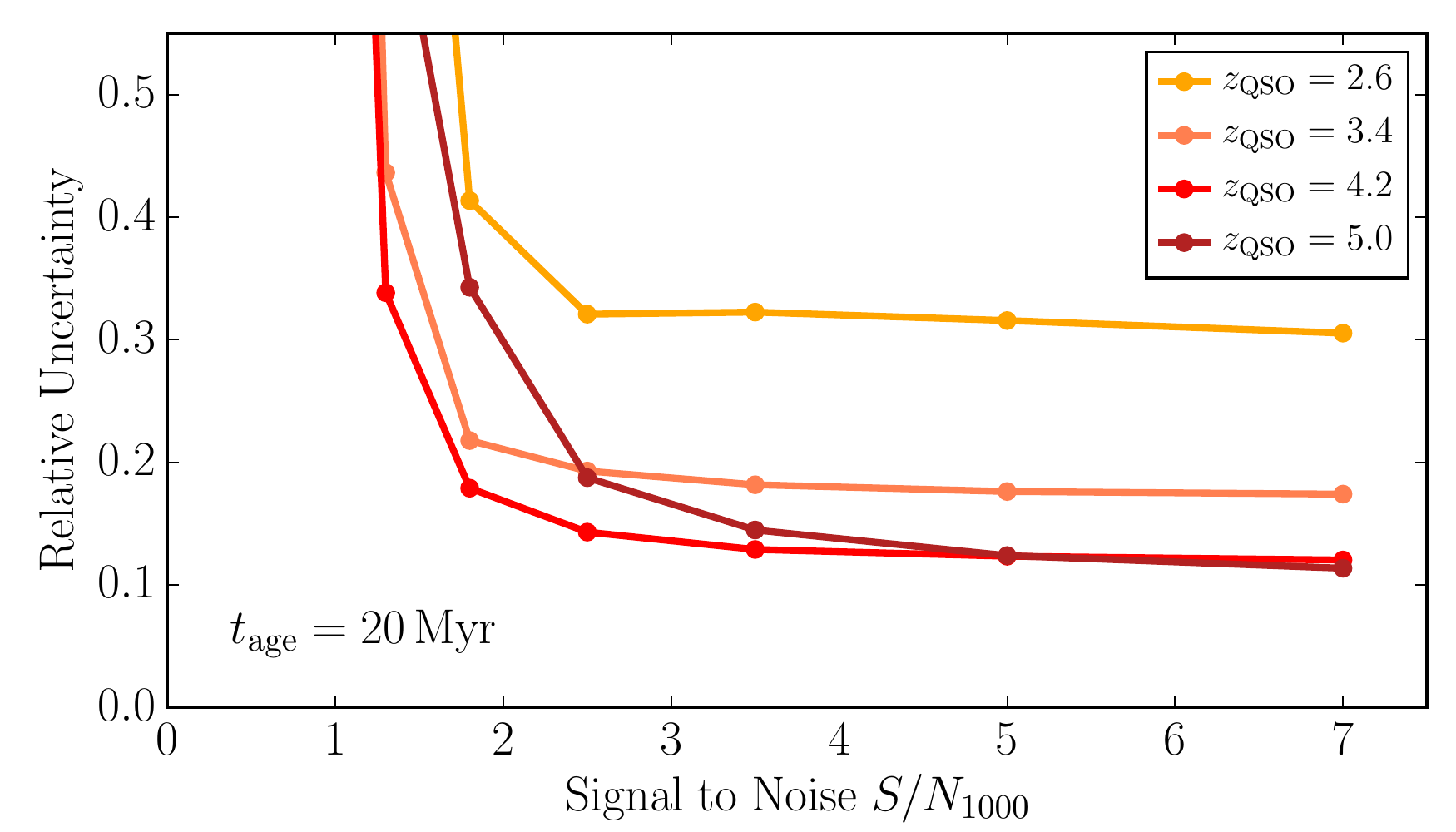}
 \includegraphics[width=\linewidth]{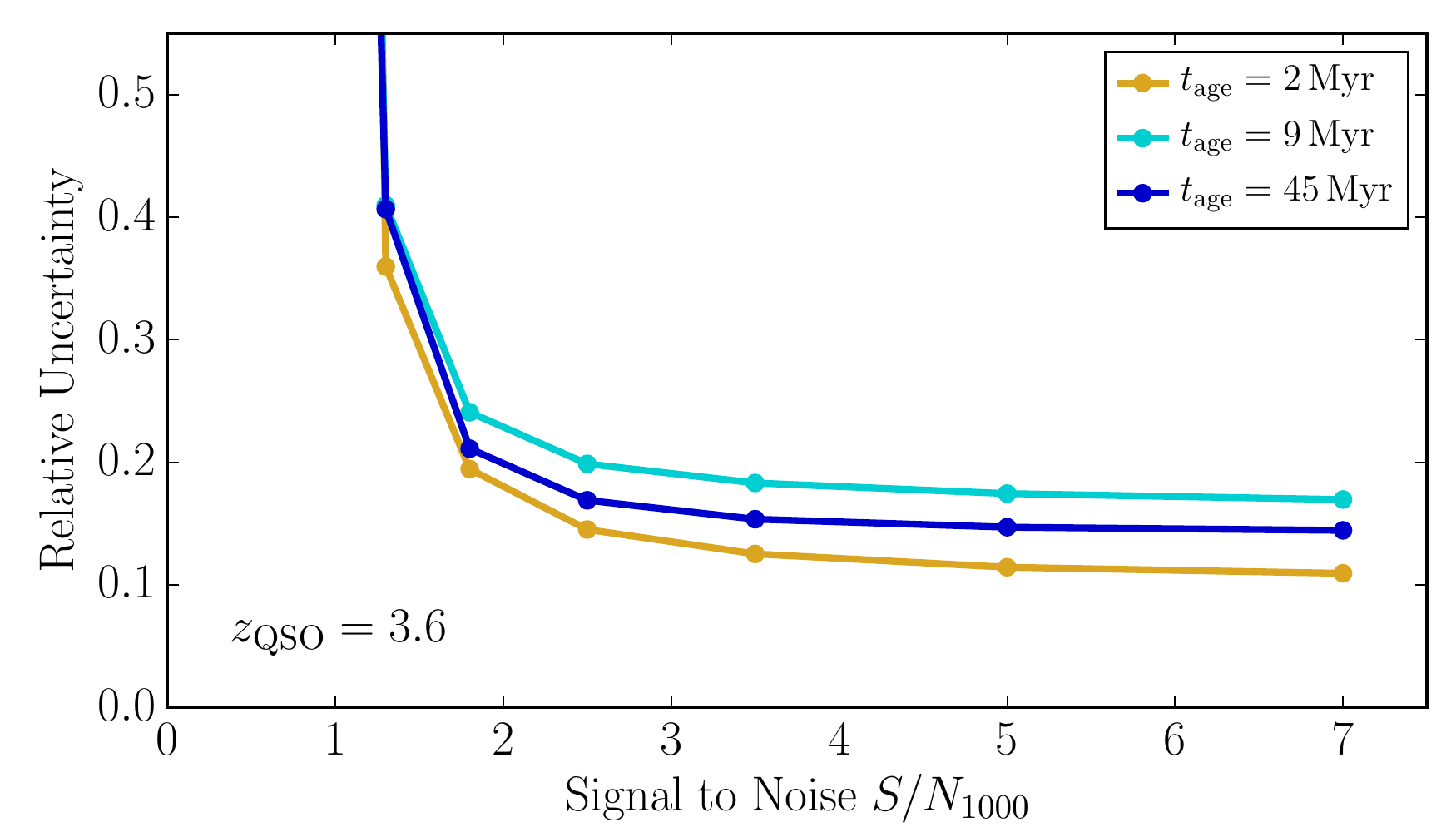}
 \caption{
 Dependence of the achieved precision on the achieved $\rm{}S/N_\mathrm{1000}$ of the data. 
 For each data quality, quasar age and quasar redshift we show the achieved precision averaged of 25 mock datasets.
 The top panel displays the behavior for four different redshifts at fixed $\tage=20\Myr$. The bototm panel shows three different quasar ages for $z_\mathrm{QSO}=3.6$.
 Further model parameters are listed in Table~\ref{Tab:Sim_Grid_z}.
 }
 \label{Fig:Sequence_SN}
\end{figure}

In \S\ref{Sec:SN_Requirement} we argued that a relatively low $\rm{}S/N$ is sufficient for our analysis since the stochastic IGM absorption causes by itself a substantial amount of noise in the transmission measurement. In this section, we now quantify this effect and determine the actual dependence of our parameter inference on the data quality.
We therefore take the models listed in Table~\ref{Tab:Sim_Grid_z} and re-compute them with different $\rm{}S/N_{1000}$. The associated continuum error is adjusted as well, following the procedure described in \S\ref{Sec:ContinuumError}.

The dependence on the achieved $\rm{}S/N_\mathrm{1000}$ shows some diversity for different quasar redshifts and quasar ages. We therefore show a broad selection of curves in Figure~\ref{Fig:Sequence_SN}. The top panel varies $\rm{}S/N$ at fixed $\tage= 20\Myr$ illustrating different redshifts; the bottom panel fixes the redshift to $z_\mathrm{QSO} = 3.6$, and varies the the quasar age. In general one sees that for $\rm{}S/N_{1000} < 2$, the achieved precision quickly deteriorates, whereas, above $\rm{}S/N_\mathrm{1000}\gtrsim 2.5$ the curves flatten indicating that the uncertainty on the inferred quasar age depends only very weakly on the data quality. Increasing the $\rm{}S/N$ of the data slightly improves the precision, but this gain is so small that it is in practice probably not worth acquiring data with $\rm{}S/N_{1000}>3.5$.

Our generally adopted value of $\rm{}S/N_\mathrm{1000}=5$ is for all $\tage$ and $z_\mathrm{QSO}$ very much on the flat region of the curves in Figure~\ref{Fig:Sequence_SN}, where the precision of the
parameter inference is not limited by the $\rm{}S/N$ of the data. It is therefore worthwhile to consider substantially relaxing the requirement on data quality e.g. to $\rm{}S/N_\mathrm{1000}=2.5$,
which might yield only slightly inferior results at approximately one fourth of the exposure time. However, data quality still has to be good enough to properly identify the objects as high-redshift background galaxies and determine their redshifts which in practice often requires $\rm{}S/N \simeq 5$ for sources which do not have strong emission lines.

\subsection{Dependence on Spectral Resolution}
\label{Sec:ResResolvingPower}

\begin{figure}
 \centering
 \includegraphics[width=\linewidth]{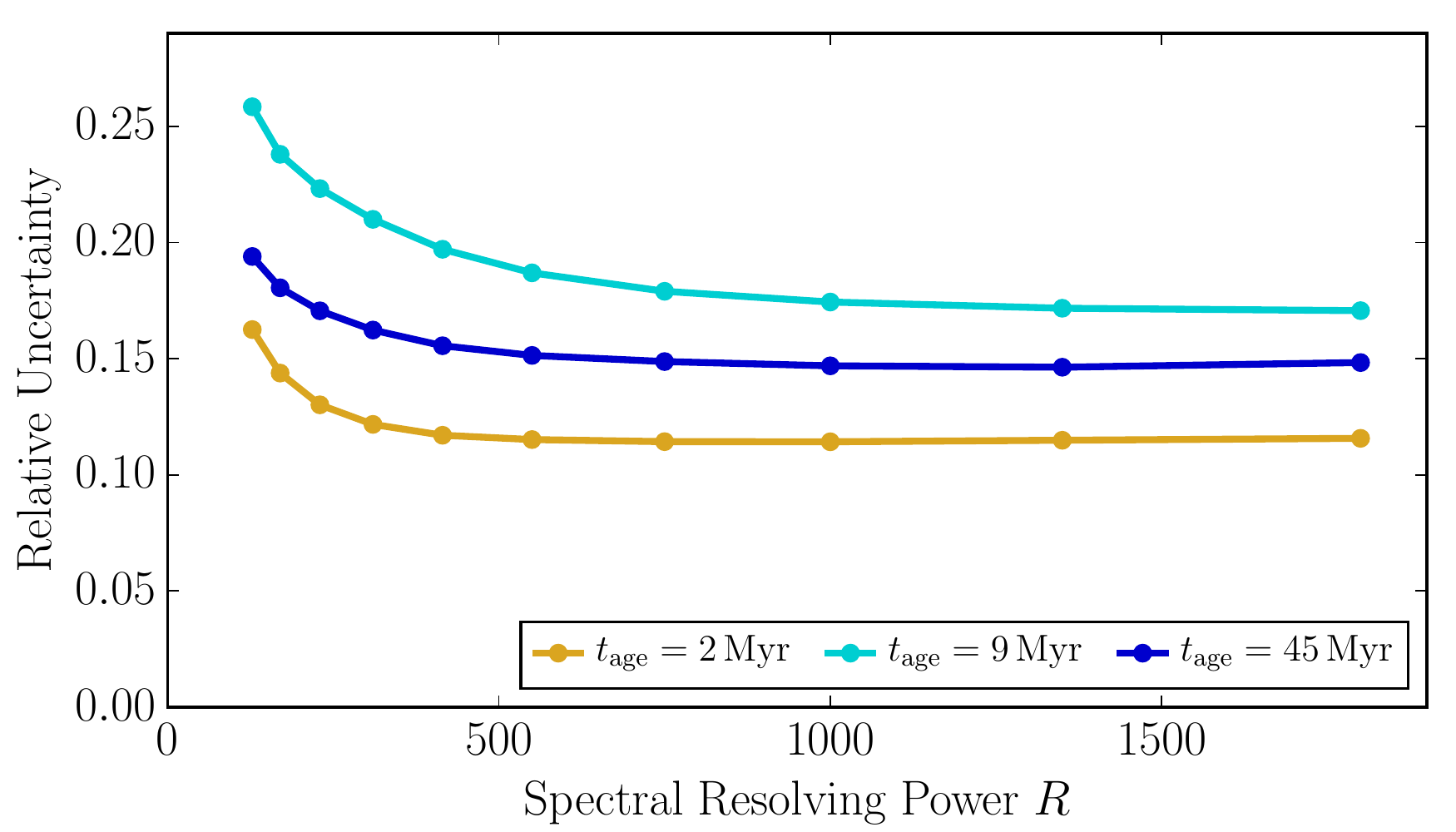}
 \caption{Dependence of the achieved precision on the spectral resolution of the data. The models are based on the $z_\mathrm{QSO}=3.6$ model listed in Table~\ref{Tab:Sim_Grid_z} and recomputed with varying spectral resolutions. For each resolving power $R$ and three different quasar ages, the average precision of 25 mock datasets is shown. The $\rm{}S/N$ per $300\,\mathrm{km\,s^{-1}}$ resolution element is kept fixed at $\rm{}S/N_{1000}=5$, corresponding to the same number of detected photons and therefore same exposure time for all shown models.
 }
 \label{Fig:Sequence_R}
\end{figure}

Another aspect of the observing strategy is the requisred spectral resolution. As argued in \S\ref{Sec:R_Requirement}, $\lya$ tomography will at some point be limited by the peculiar velocities in the IGM and in \S\ref{Sec:SimulatedData} we determined that these redshift space distortions have an amplitude of $\approx${}\,$300\,\mathrm{km\,s^{-1}}$, indicating that spectral resolving powers of the order of $R\approx1000$ should be sufficient. In Figure~\ref{Fig:Sequence_R} we show that the actual requirement is even lower. Provided $R\geq750$, we observe an extremely weak dependence of the resulting age precision on  the resolution. Even below this, the achieved precision is only moderately impacted. 

Therefore, spectral resolution is essentially of no concern for characterizing quasar light echoes. Nearly every multi-object spectrograph should deliver sufficient resolution, even in low
resolution modes.  Resolving powers of $R\lesssim200$ already come close to the regime of slitless grism or prism spectroscopy. It might be possible to benefit from these observing strategies for $\lya$ forest tomography without resulting in a significant penalty on the achieved precision\footnote{We stress however, that slitless spectroscopy suffers from substantially higher sky noise compared to   slit spectroscopy. If at all, slitless spectroscopy is therefore only an option for space based observations.}. However, the adopted spectral resolution has to be good enough to identify the objects as high-redshift background galaxies. In any case, our adopted fiducial resolving power of $R=1000$ is more than sufficient.

\subsection{ Dependence on Sightline Density }
\label{Sec:ResSightlineDensity}

A further key parameter for the tomographic mapping of quasar light echoes is the density of background sightlines. Clearly, this depends on the limiting magnitude of the observations and the redshift of the foreground quasar. The exact relations between these parameters is illustrated in Figure~\ref{Fig:SightlineDensity}.

\begin{deluxetable}{ccccccccc}
\centering
\tablecolumns{8}
\tablewidth{0pc}
\tablecaption{Parameter of Simulations used in \S\ref{Sec:ResSightlineDensity}}
\tablehead{
\colhead{$z_\mathrm{QSO}$}	& \colhead{$M_{1450}$}	& \colhead{$r_\mathrm{lim}$}	& \colhead{$D_\mathrm{SL}$}	& \colhead{$\mathrm{FoV}$}	& \colhead{$N_\mathrm{SL}$}	&	\colhead{$R$}	&	\colhead{$\rm{}S/N_{1000}$}	\\
\colhead{}			& \colhead{$\Mag$}	& \colhead{$\Mag$}		& \colhead{$\cMpc$}		& \colhead{$'$}		& \colhead{}			&	\colhead{}	&	\colhead{}	
}
\startdata
$3.6$   & $-28.7$       & $25.0$        & $3.36$        & $16$  & $ 60$         & $1000$         & $5.0$         \\ 
$3.6$   & $-28.7$       & $24.6$        & $4.57$        & $16$  & $ 36$         & $1000$         & $5.0$         \\ 
$3.6$   & $-28.7$       & $24.2$        & $6.57$        & $16$  & $ 18$         & $1000$         & $5.0$         \\ 
$3.6$   & $-28.7$       & $23.8$        & $10.18$       & $16$  & $  6$         & $1000$         & $5.0$         
\enddata
\label{Tab:Sim_Grid_r}
\end{deluxetable}

To explore how our precision depends on the sightline density, we compute another set of models for which we vary the limiting magnitude of the observations. This atlers the average separation of sightlines and, since we keep the FoV constant, the number of sightlines. For simplicity, we only conduct this exercise for a quasar redshift of $z_\mathrm{QSO}=3.6$. The exact details of the models used for this are given in Table~\ref{Tab:Sim_Grid_r}.

\begin{figure}
 \centering
 \includegraphics[width=\linewidth]{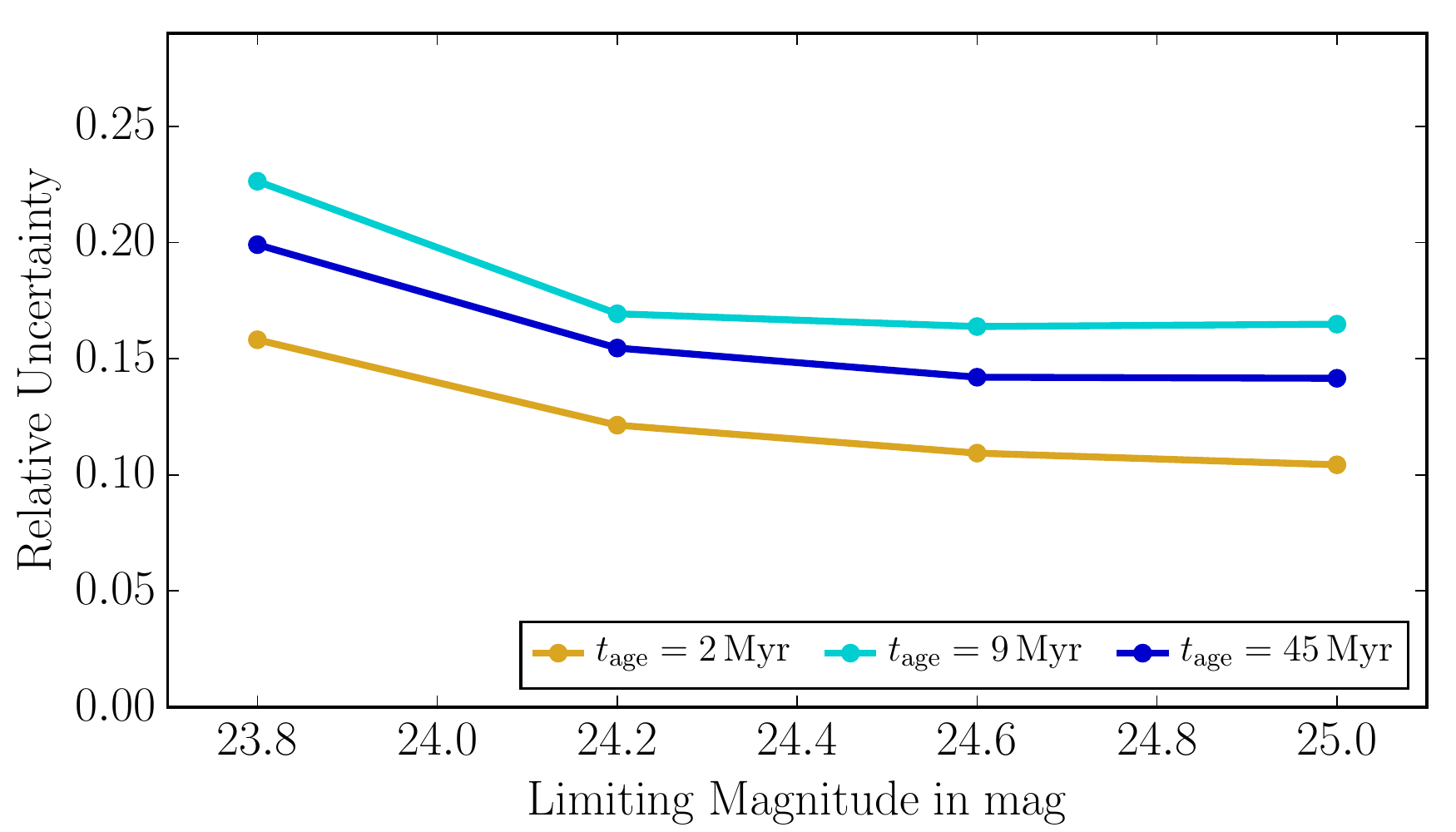}
 \caption{Dependence of the achieved precision on the limiting magnitude of the observations and therefore on the background sightline density. The curves are based on the $z_\mathrm{QSO}=3.6$ model listed in Table~\ref{Tab:Sim_Grid_z} and recomputed for different limiting magnitudes $23.8\Mag \leq r_\mathrm{lim} \leq 25.0\Mag$. This results in average sightline separations $D_\mathrm{SL}$ between $10\cMpc$ and $3.4\cMpc$. Since the FoV is fixed at $16'$, the number of background sightlines $N_\mathrm{Sl}$ varies between 6 and 60. See also Table~\ref{Tab:Sim_Grid_r} for details of the models.
 }
 \label{Fig:Sequence_rlim}
\end{figure}

The results of this analysis are shown in  Figure~\ref{Fig:Sequence_rlim}. As one can see, there is no strong dependence of the precision on the sightline density (or the limiting magnitude). Apparently, sampling the proximity zone with six background sightlines ($r_\mathrm{lim}=23.8\Mag$) already gives a reasonable estimate of the quasar age. Increasing the sightline density to yield 18 background sightlines ($r_\mathrm{lim}=24.2\Mag$) reduces the uncertainty by $\approx\,$20$\,\%$.  Any further increase yields only a marginal improvement. The strongest effect is seen for short quasar ages. Here, the region illuminated by the quasar is the smallest (see Figure~\ref{Fig:HI_TPE_Slice}) and a finer sampling by background sightlines leads to the biggest improvement.  In general,
Figure~\ref{Fig:Sequence_rlim} indicates that a survey somewhat shallower than our fiducial $r_\mathrm{lim}=24.7\Mag$ might be sufficient to constrain quasar ages.

However, we must stress that we so far consider only a highly simplified quasar emission model with $\tage$ as the only free parameter. In reality, quasar UV emission is expected to be anisotropic
\citep{Antonucci1993, Urri1995, Netzer2015}. A proper treatment of this would require a model that includes obscuration and orientation effects. Constraining all four parameter that specify such a model would probably require a larger number of background sightlines and smaller sightline separations.  For our current isotropic emission model, even with relatively coarse  background sightlines sampling, the size of the illuminated region and therefore $\tage$ can still easily  be inferred from locations in front of the quasar ($R_\perp \ll 0$, $z < z_\mathrm{QSO}$, left in Figure~\ref{Fig:HI_TPE_Slice}) where these sightlines intersect the parabolic shaped illuminated region. This can also be seen in Figure~\ref{Fig:TimeRetardation}, where even sightlines with $R_\perp=16\cMpc$ probe time differences shorter than $\Delta{}t<2\Myr$.  The situation will be different if the emission geometry of the quasar must also be determined. In that case, an average sightline separation comparable or smaller than the age of the quasar, i.e. $D_\mathrm{SL} < \mathrm{c} \; \tage$, are likely to be required. We will characterize the exact requirements for this more complex case in detail in a future paper.

We also stress that throughout our analysis in this section we used the brightest quasar at any given redshift (see Figure~\ref{Fig:BrightestQuasars}). When using fainter quasars, the proximity zone is less extended and we expect that in return one requires a denser packed background sightline pattern to achieve the same precision.

\subsection{ Dependence on Field-of-View}
\label{Sec:ResFoV}

\begin{deluxetable}{ccccccccc}
\centering
\tablecolumns{8}
\tablewidth{0pc}
\tablecaption{Parameter of Simulations used in \S\ref{Sec:ResFoV}}
\tablehead{
\colhead{$z_\mathrm{QSO}$}	& \colhead{$M_{1450}$}	& \colhead{$r_\mathrm{lim}$}	& \colhead{$D_\mathrm{SL}$}	& \colhead{$\mathrm{FoV}$}	& \colhead{$N_\mathrm{SL}$}	&	\colhead{$R$}	&	\colhead{$\rm{}S/N_{1000}$}	\\
\colhead{}			& \colhead{$\Mag$}	& \colhead{$\Mag$}		& \colhead{$\cMpc$}		& \colhead{$'$}		& \colhead{}			&	\colhead{}	&	\colhead{}	
}
\startdata
$3.6$   & $-28.7$       & $24.2$        & $6.57$        & $ 7$  & $  6$         & $1000$         & $5.0$         \\ 
$3.6$   & $-28.7$       & $24.2$        & $6.57$        & $14$  & $ 18$         & $1000$         & $5.0$         \\ 
$3.6$   & $-28.7$       & $24.2$        & $6.57$        & $21$  & $ 36$         & $1000$         & $5.0$         \\ 
$3.6$   & $-28.7$       & $24.2$        & $6.57$        & $28$  & $ 60$         & $1000$         & $5.0$         \\ 
$3.6$   & $-28.7$       & $24.2$        & $6.57$        & $35$  & $ 90$         & $1000$         & $5.0$         \\ 
$3.6$   & $-28.7$       & $24.2$        & $6.57$        & $42$  & $126$         & $1000$         & $5.0$          
\enddata
\label{Tab:Sim_Grid_F}
\end{deluxetable}

\begin{figure}
 \centering
 \includegraphics[width=\linewidth]{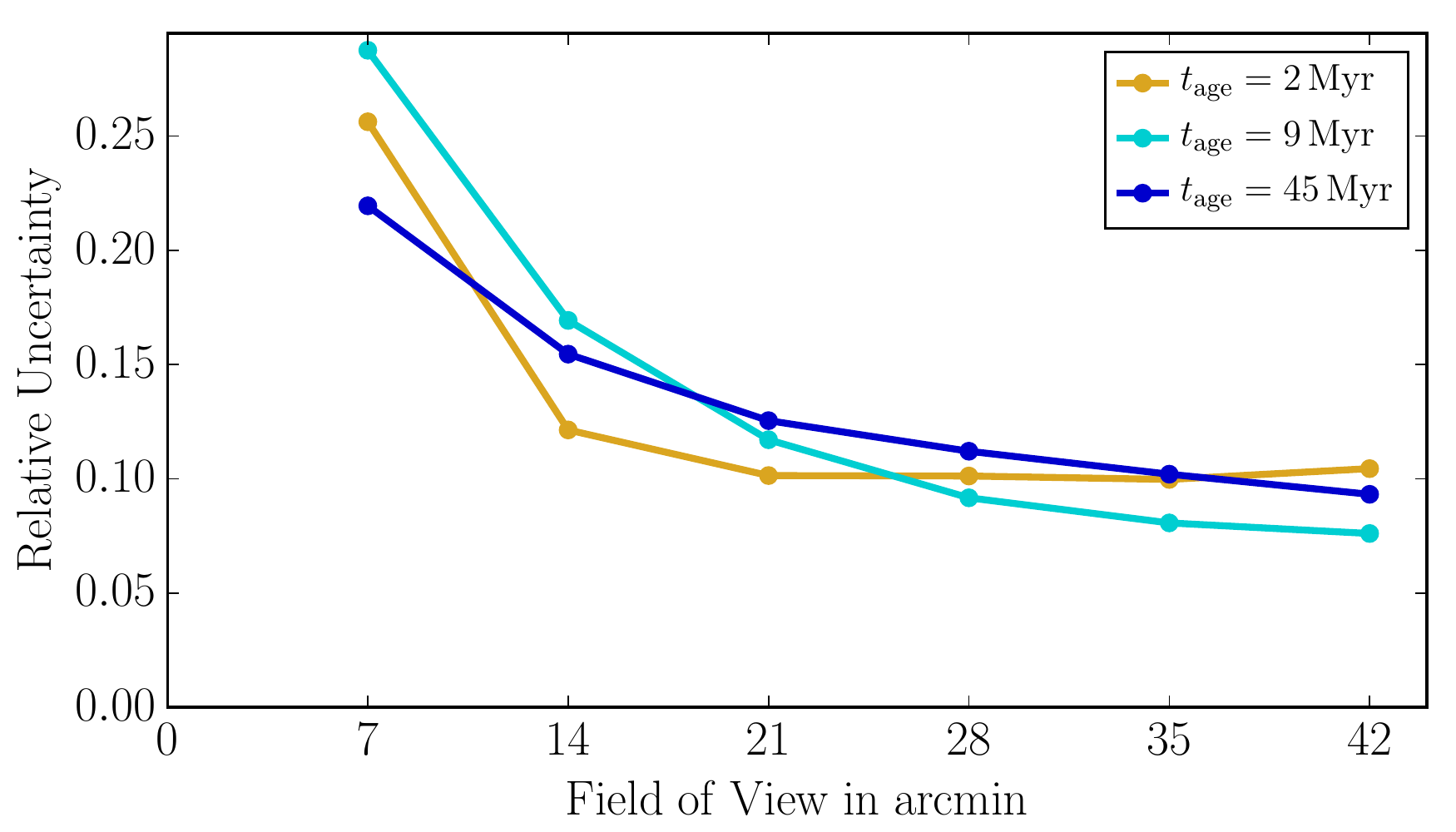}
 \caption{Dependence of the achieved precision on the field-of-view covered by the observations. The curves are based on the $r_\mathrm{lim} = 24.2\Mag$ model listed in Table~\ref{Tab:Sim_Grid_r} and recomputed for different fields between $7'$ and $35'$ diameter and therefore different number of background sightlines. The average sightline separations is kept fixed at $D_\mathrm{SL} = 6.6\cMpc$. See also Table~\ref{Tab:Sim_Grid_F} for details of the models.
 }
 \label{Fig:Sequence_FoV}
\end{figure}

Similar to the limiting magnitude of the observations, the achievable precision
is influenced by the field-of-view covered by the observations. Both aspects together define the number of sightlines that sample the proximity zone.
We compute a set of models at $z_\mathrm{QSO}=3.6$ with fixed $r_\mathrm{lim} = 24.2\Mag$ and therefore fixed $D_\mathrm{SL} = 6.6\cMpc$, but vary the diameter of the field between $7'$ and $35'$. Details of the models are listed in Table~\ref{Tab:Sim_Grid_F} and the results are shown in Figure~\ref{Fig:Sequence_FoV}.

Again, given the relatively high $z_\mathrm{QSO}$ and simple single parameter quasar emission model, a small number of background sightlines is sufficient to constrain the model parameter $\tage$. However, increasing the FoV and therefore the number of background sightlines does improve the precision substantially from $\approx20\,\%$ at $7'$ FoV to better than $10\,\%$ at $42'$. The dependence is different for different quasar ages. While for $\tage=2\Myr$ we find no significant improvement, longer quasar ages and in particular intermediate ages around $\sim10^7\yr$, which usually have the largest uncertainties, gain the most from a larger FoV. 

Our understanding is that the precision of the analysis is limited primarily by redshift space distortions. Probing a larger field helps to average these down. Since redshift space distortions are coherent over large length scales, this can only be done by actually probing a larger volume but not by increased sightline density in a confined volume. A larger field-of-view is therefore clearly beneficial to our provided the quasar ages are sufficiently long that their proximity zones extend beyond the adopted FoV.
However, most currently existing multi-object spectrographs have a rather limited FoV $<10'$, requiring multiple pointings to cover a large field. The gain in precision is therefore probably not big enough to justify the substantial increase in observing time required to map the full extent of the proximity region. The exception is the Subaru Prime Focus Spectrograph which will have a field-of-view $78'$ in diameter. This new instrument will be able to cover the full proximity zone and deliver spectra of a large number of background sightlines in a single pointing, highly beneficial for our
application.

\subsection{Dependence on Quasar Luminosity}

\begin{figure}
 \centering
 \includegraphics[width=\linewidth]{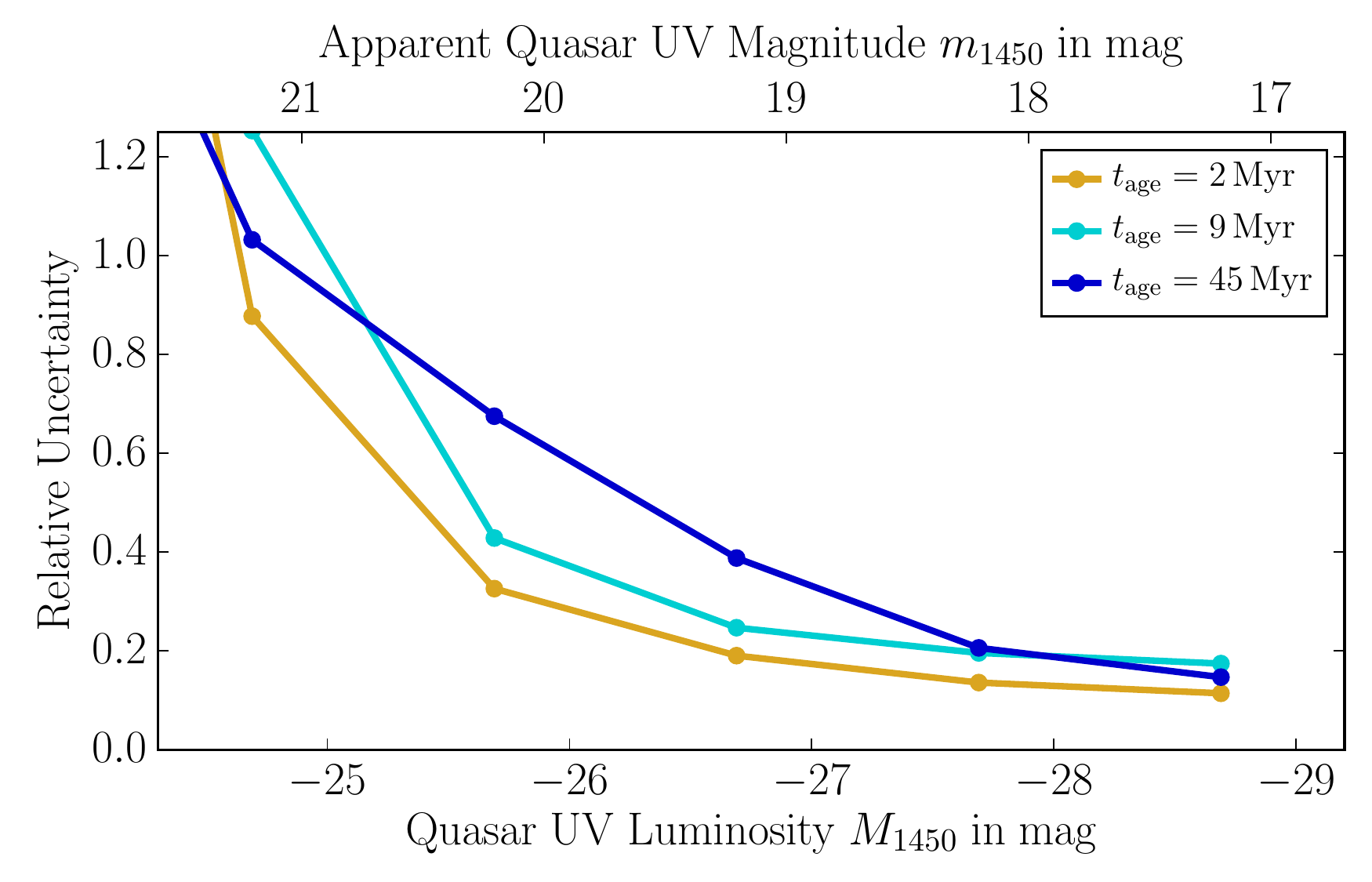}
 \caption{Dependence of the achieved precision on the luminosity of the foreground quasar. The curves are based on the $z_\mathrm{QSO}=3.6$ model listed in Table~\ref{Tab:Sim_Grid_z} and recomputed for different values of $M_{1450}$. For each quasar luminosity and three quasar ages, the average precision of the 25 mock datasets is shown.
 }
 \label{Fig:Sequence_M1450}
\end{figure}

The luminosity of the foreground quasar is clearly another crucial factor for our analysis. Brighter quasars have larger proximity zones that can be probed by more background sightlines. Even in cases where the proximity zone extends beyond the FoV and the number of background sightlines is limited by the instrument rather than the quasar, the proximity zone of brighter quasars has a larger extent along the line-of-sight. Also, for a given separation from the foreground quasar, a more luminous quasar will cause a stronger IGM transmission enhancement, even though the transmission is anyway close to unity in a substantial fraction of the illuminated volume (see Figure~\ref{Fig:HI_FluxStatistic}). It is therefore always beneficial to target the brightest quasars.

In Figure~\ref{Fig:Sequence_M1450} we show the dependence of age
precision on quasar luminosity. Clearly, the highest precision is
achieved for the brightest quasars and the precision steadily
decreases for fainter quasars. When comparing to
Figure~\ref{Fig:GammaQSO}, one can see that a quasar with
$M_{1450}=-29\Mag$ exceeds the UV background up to an angular distance
of $30'$ and therefore far beyond our fiducial FoV which has a $8'$ radius. The
proximity zone size scales as the square root of the quasar
luminosity. One can therefore estimate at which luminosity the quasar and
UV background contribute equally to the \ion{H}{i} photoionization
rate at the edge of the FoV.  This happens for
$M_{1450}\approx-26.2$. Moving from the brightest quasars
down to this luminosity, one observes that quasar age precision
deteriorates relatively slowly.  For fainter quasars however, the outermost background sightlines do
not probe the proximity zone anymore, leading to a rapid loss of
precision. 

In general, mapping quasar light echoes with $\lya$ forest tomography works best for the absolutely brightest quasars. However, also quasars up to break magnitude of the quasar luminosity function $M_{1450}^* \approx -27.3\Mag$ \citep{Kulkarni2018}, are reasonable targets for which quasar ages could be deduced with only slightly reduced precision. Therefore, as can be seen in Figure~\ref{Fig:BrightestQuasars}, many quasars ($\gg100$) are available as potential targets which offers, at least in principle, the opportunity to apply $\lya$ forest tomography to a substantial number of quasars and to study the distribution of lifetimes.

\subsection{Impact of Continuum Uncertainties}

\begin{figure}
 \centering
 \includegraphics[width=\linewidth]{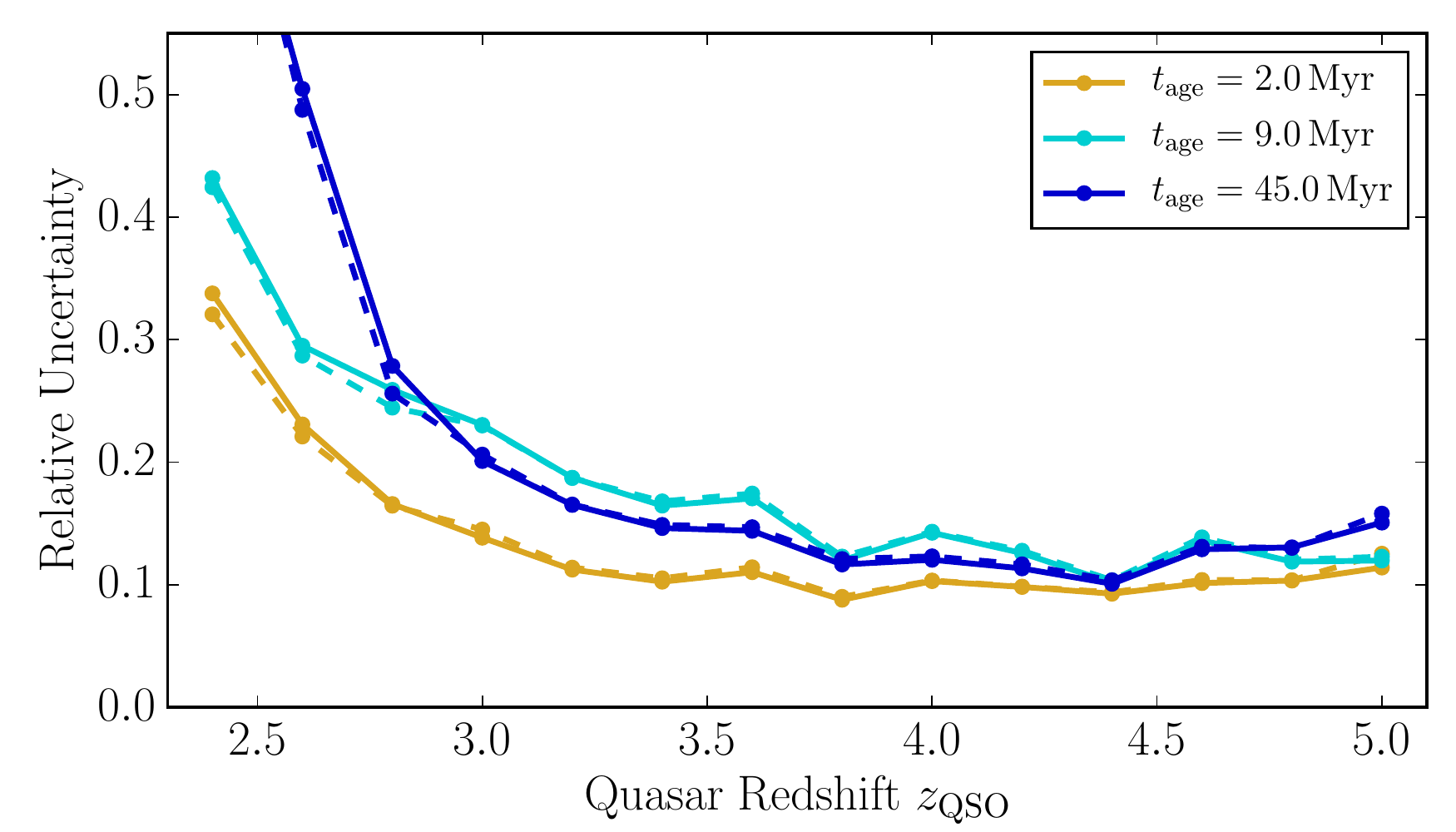}
 \caption{Dependence of the achieved precision on the adopted continuum uncertainty. The plot is similar to Figure~\ref{Fig:Sequence_zQSO} and shows for each quasar redshift and three different quasar ages the average precision of 25 mock datasets. However, we show as dashed lines the achieved precision when using the usual scheme for continuum uncertainties (see \S\ref{Sec:ContinuumError}) and as solid lines the case with perfect continuum fits.
 }
 \label{Fig:Sequence_zQSO_C050}
\end{figure}

Continuum uncertainties could in principle have a substantial impact on our analysis. All plots shown above do include our standard scheme for modeling continuum uncertainties as described in \S\ref{Sec:ContinuumError}. However, it is also worthwhile to understand how much continuum uncertainties are degrading our precision relative to the ideal case of no continuum errors. We therefore re-run the full analysis procedure, with model parameters identical to the ones listed in Table~\ref{Tab:Sim_Grid_z}, but assuming perfect knowledge of the continuum. The result is shown in Figure~\ref{Fig:Sequence_zQSO_C050}.

Clearly, continuum uncertainties at level indicated by Equations~\ref{Eq:Noise} and \ref{Eq:Noise_sigma2} ($10\,\%$ for $\rm{}S/N_{1000}=5$) have a negligible impact on our results. This slightly changes when considering poorer data quality. For example, for $\rm{}S/N_{1000}=2.5$ ($18\,\%$ continuum error), we find a noticeable deviation between the fit including the continuum uncertainty and a perfect continuum. However, the effect is rather small and the uncertainty in the $\tage$ estimate increases only from e.g. 20\,\% to 23\,\%, emphasizing that continuum errors are no major concern in the context of our study.

\section{Conclusion}

In this paper we presented a novel method to map quasar light echoes and infer quasar ages, employing $\lya$ forest tomography.  The method utilizes the $\lya$ forest absorption in the spectra of faint background galaxies ($m_\mathrm{r}\approx24.7\Mag$) to probe the ionization state of the intergalactic medium in the vicinity of bright foreground quasars.  The UV radiation of quasars has a strong impact on the IGM and can substantially enhance the ionization state of the gas, resulting in enhanced IGM transmission which is known as \textit{proximity effect}. Relying on faint galaxies as background sources results in a high sightline density (1000 per degree) and allows one to probe the proximity zone of individual foreground quasars with many (10 to 100) background sightlines. These detailed observations allow one to construct a three-dimensional map of the quasar light echo and ultimately to constrain the quasars emission history and emission geometry.

In this study, we developed a full end-to-end simulation pipeline to model this experiment and we demonstrated that it is feasible on current 8m class telescopes. In this context, we described a
collection of observational parameters (quasar luminosities, sightline densities, etc) which set the framework for future tomographic observations of light echoes.  (\S\ref{Sec:ObservationalSetup}). We then constructed a suite of models of the $\lya$ transmission in the quasar proximity region, based on \Nyx{} cosmological hydrodynamical simulations which were postprocessed with a photoionization model (\S\ref{Sec:HydroSims}, \ref{Sec:SimulatedData}). We introduced a novel \emph{likelihood-free} Bayesian analysis formalism (\S\ref{Sec:LikelihoodComputation}) which enables statistical comparison of IGM tomography observations to these models, which delivers robust posterior probability distributions of the model the model parameter ($\tage$), fully accounting for the strong
correlations in the tomographic map and the non-Gaussian nature of the IGM transmission. We thoroughly tested this new machinery on mock observations (\S\ref{Sec:Results}) which leads to the following
conclusions:

\begin{itemize}
\item IGM tomography observations of quasar light echoes are capable of yielding precise unbiased constraints on the quasar emission history. The achievable relative precision on quasar age assuming realistic observing times ($10\,\mathrm{ks}$) using existing instruments on 8-10\,m class telescopes is $\approx20\,\%$.
\item The highest relative precision (10\,\%) is achieved for very short ($1\Myr$) or very long ($100\Myr$) quasar ages, while for intermediate $\tage\sim10\Myr$ we can measure $\tage$ to about 20\,\% precision.
\item Our new method delivers satisfactory ($<25\,\%$) constraints for all quasar redshifts $3 < z_\mathrm{QSO} <5$ with weak dependence of the precision on redshift.
\item A spectral resolution as low as $R=750$ is completely sufficient. Using even lower resolution down to $R\approx200$ might be possible without a significant loss in precision.
\item The minimal required signal to noise ratio per $300\,\mathrm{km\,s^{-1}}$ bin is $\rm{}S/N_{1000}\approx2.5$. Higher ${\rm S\slash N}$ data does improve the precision but only slightly.
\item The brightest quasars are the best targets. However, quasars as  faint as $M_{1450} < -27.5\Mag$ can be used with little loss of precision. This implies that $\gg$\,100 targets are available at    $3<z_\mathrm{QSO} < 5$ opening up the possibility of mapping out the distribution of quasar ages for statistical samples.
\end{itemize}

This demonstrates that $\lya$ forest tomography has the potential to measure the emission properties of individual bright quasars and in particular constrain the age of quasars in the range from $1\Myr$ up to $100\Myr$ with $\approx20\,\%$ precision.
While we focused in this paper on the general feasibility of the method, the observational requirements, and the ability to constrain the quasar age, we will in upcoming papers investigate more complex models for the quasar emission, like a more realistic lightcurve and non isotropic emission, which involves a straightforward generalization of the modeling and statistical framework presented here. In particular, we intend to infer for individual quasars their orientation and obscuration geometry which is so far even in a statistical sense only poorly constrained \citep{Brusa2010, Assef2013, Lusso2013, Marchesi2016}. Such measurements will then allow to test quasar unification models \citep{Antonucci1993, Urri1995, Netzer2015} and compare them to other explanations for quasar obscuration \citep[e.g.][]{Elvis2000,  Elitzur2006, Honig2017}. There might also be synergies with other quasar lifetime measurements, in particular the ones derived from the \ion{He}{ii} line-of-sight proximity effect. Some of the quasars for which \citet{Khrykin2018} have recently presented constraints are viable targets for our $\lya$ forest tomography. This offers the opportunity for a cross-check between two rather different methods. 

We showed that the current generation of instruments on 8-10\,m class telescopes are capable of deriving meaningful constraint on quasar emission properties via $\lya$ forest tomography.  However, the
introduction of new highly-multiplexed multi-object spectroscopic facilities on $8-10\,\mathrm{m}$ telescopes in the near future, in particular the Subaru Prime Focus Spectrograph, will give a tremendous boost to tomographic observations like the one described in this
work. This will make $\lya$ forest tomography one of the key techniques to study the emission histories and emission geometries of quasars and deliver unprecedented insight into quasar activity cycles, quasar physics, and the buildup of supermassive black holes.

\section*{Acknowledgments}
We are grateful to Frederick Davies and Michael Walther for many useful discussions about statistical
and numerical methods. We would also like to thank the members of the ENIGMA\footnote{\url{http://enigma.physics.ucsb.edu/}} group at the Max Planck Institute for Astronomy (MPIA) and UCSB for constructive
comments on an early version of this manuscript. 
\bibliographystyle{APJ}
\bibliography{Literature}

\end{document}